\pdfoutput=1
\documentclass[nolongbibliography,twocolumn,superscriptaddress,prb,showpacs,showkeys]{revtex4-2}

\usepackage[utf8]{inputenc} 

\usepackage{graphicx} 
\usepackage{bm}
\usepackage{bbold}
\usepackage{amsmath}

\begin{document}
\title{Core-hole excitations using the projector augmented-wave method and the Bethe-Salpeter equation}
\author{Martin Unzog}
\email{martin.unzog@univie.ac.at}
\affiliation{%
University of Vienna, Faculty of Physics \& Computational Materials Physics \& Vienna Doctoral School in Physics,  Boltzmanngasse 5, 1090 Vienna, Austria
}%

\author{Alexey Tal}
\email{alexey.tal@vasp.at}
\affiliation{%
VASP Software GmbH, Sensengasse 8/12, 1090, Vienna, Austria 
}%

\author{Georg Kresse}%
\email{georg.kresse@univie.ac.at}
\affiliation{%
University of Vienna, Faculty of Physics, Computational Materials Physics, Kolingasse 14-16, 1090 Vienna, Austria
 }%
\begin{abstract}
We present an implementation of the Bethe-Salpeter equation (BSE) for core-conduction band pairs within the framework of the projector augmented-wave method. For validation, the method is applied to the $K$-edges of diamond, graphite, hexagonal boron-nitride, as well as four lithium-halides (LiF, LiCl, LiI, LiBr). We compare our results with experiment, previous theoretical BSE results, and the density functional theory-based supercell core-hole method. In all considered cases, the agreement with experiment is excellent, in particular for the position of the peaks as well as the fine structure. Comparing BSE to supercell core-hole spectra we find that the latter often qualitatively reproduces the experimental spectrum, however, it sometimes lacks important details. This is shown for the $K$-edges of diamond and nitrogen in hexagonal boron-nitride, where we are capable to resolve within the BSE experimental features that are lacking in the core-hole method. Additionally, we show that in certain systems the supercell core-hole method performs better if the excited electron is added to the background charge. We attribute this improved performance to a reduced self-interaction.

\end{abstract}

\maketitle

\section{Introduction}

X-ray Absorption Spectroscopy (XAS) experiments probe the transition probability of exciting an electron from a core state to a conduction band state. XAS spectra furnish information about the local chemical environment of the probed atom, e.g.\ coordination number or oxidation state \cite{Yano2009}. However, experimental XAS spectra are difficult to interpret without theoretical input. 

To simulate XAS spectra, two fairly simple \emph{ab initio} approaches exists. In the initial-state approximation the XAS spectrum is calculated via Fermi's golden rule, where the initial state is a core state and the final state a conduction band state. In the final-state approximation the final-state wavefunction is calculated self-consistently using Kohn-Sham Density Functional Theory (DFT) \cite{Hohenberg1964} by removing one core electron and placing it in the conduction bands. 

These methods have a couple of drawbacks. In the initial-state approach, electrons and holes are independent particles and any interactions between them are neglected. In the final-state approach, supercells are required to avoid spurious interactions between core-holes located in different unit cells, which is why this approach is here called Supercell Core-Hole (SCH) method. Further, one needs to assume that the approximate density functionals remain valid for excited states. Finally, the excited electron is usually placed into the conduction band edge, and it is assumed that the resulting renormalized one-electron energies for all conduction bands will be a good approximation for the XAS fine structure. 

By looking at these restrictions it is clear that two aspects need to be considered in the simulation of XAS spectra. First, one needs an accurate description of the electronic structure, ideally including many-body correlations for the interaction of electrons and holes. Secondly, the use of supercells should be avoided.

The state of the art approach that resolves both issues is the $GW$+BSE method. It combines \emph{ab initio} ground state electronic structure calculations with techniques of Many Body Perturbation Theory (MBPT) \cite{Onida2002}. By calculating the one-particle Green's function $G$ various one-body properties can be obtained, e.g.\ quasiparticle band gaps and quasiparticle energies \cite{Hybertsen1986}. 

For a description of the mutual interaction of the electrons and holes one needs to go beyond one-particle theory and solve the equation of motion for the two-particle Green's function -- the Bethe-Salpeter Equation (BSE) \cite{Sham1966,Martin2016}. The BSE is an integral equation that can be recast into an eigenvalue equation, where exchange and the screened attraction of electron and hole are incorporated in the resulting BSE matrix. The BSE approach has first been applied to optical spectra of semiconductors in the late seventies \cite{Hanke1979}, \emph{ab initio} based approaches to the BSE followed in the early 2000s  \cite{Benedict1998,Albrecht1998,Horst1999,Rohlfing2000}.

In these approaches, the orbitals are usually obtained from a standard ground state DFT calculation. In the second step, the  Green's function is calculated using the DFT orbitals and one-electron energies. Many-body correlation effects are included  by replacing the Kohn-Sham exchange-correlation potential by the self-energy in the $GW$ approximation and solving the Dyson equation for the interacting Green's function. The one-electron energies are then updated by equating them to the poles of the resulting interacting Green's function. In most cases, a single iteration is performed. Finally, a simplified two-particle BSE is solved. Often one refers to this approach as $GW$+BSE. 

The $GW$+BSE approach has been extended from the optical to the X-ray regime by covering excitations from core levels to the conduction band. The earliest implementation was presented by Shirley \emph{et al.}\ \cite{Shirley1998,Soininen2001} within a pseudopotential scheme. This work culminated in the OCEAN package \cite{Vinson2011,Gilmore2015,Shirley2020}. Further adaptions within all-electron full-potential methods were established by Olovsson \emph{et al.}\ in \texttt{exciting} \cite{Olovsson2009}  and subsequently by Laskowski \emph{et al.}\ in \textsc{WIEN2K}  \cite{Laskowski2010}. To the best of our knowledge, most available electronic structure codes that implement the Projector Augmented-Wave (PAW) methodology calculate XAS spectra via the supercell core-hole method \cite{Gao2009,Gougoussis2009,Mazevet2010,Bunau2013,Prentice2020}.

In this work, we present an implementation of the $GW$+BSE scheme in the PAW method for X-ray absorption spectra. We derive the BSE in the PAW scheme and  show what modifications have to be made for core states. Our implementation is tested by calculating XAS spectra for the $K$-edges of one prototypical covalent system (diamond), two 2D van-der-Waals layered materials (graphite and hexagonal boron-nitride), and four ionic materials with shallow core states (the lithium-halides LiF, LiCl, LiBr, and LiI). We benchmark our results against  spectra obtained either by a pseudopotential or an all-electron implementation of the BSE and point out where the $GW$+BSE implementation within PAW shows improvements. We furthermore compare our results with experiment and a previously presented implementation of the SCH method \cite{Karsai2018}. 

Finally, we show that previously  obtained SCH spectra of the lithium-halides can be improved by adding the electron to the background charge instead of the lowest conduction band. We argue that the improvement comes from a reduction of self-interaction errors and connect this explanation to the $GW$+BSE approach by investigating how the spectra change if the exchange term of the BSE is omitted.

The necessary theory is presented in Section \ref{sec:theory}. In Subsection \ref{sec:BSE} we shortly summarize the BSE as well as the basics of the PAW formalism. In this subsection we will also point out the approximations made in the current implementation. In Subsection \ref{sec:matelements} we present details concerning the implementation of the BSE in the PAW formalism. The expressions for the transition probablities and the dielectric function are then found in Subsection \ref{methods_dielectric}. In Section \ref{sec:methods} we summarize the computational methods, followed by a presentation of the results in Section \ref{sec:results}. In Section \ref{sec:error} we present the improved SCH spectra for the lithium-halides and the discussion of the suggested self-interaction effect. Finally, we summarize and conclude the paper in Section \ref{sec:summary}.

\raggedbottom

\section{Theory}\label{sec:theory}
\subsection{Bethe-Salpeter equation}\label{sec:BSE}

We use notations consistent with a previous publication \cite{Sander2015}, in which further details on the implementation of the BSE in PAW can be found. We use the commonly used notation for space and time variables: $1=\{\bm{r}_1,t_1\}$, etc. In this work we treat nonmagnetic systems, however, the equations can be easily generalized to include spin indices.

The BSE is the Dyson equation for the two-particle correlation function $L$
\begin{equation}\label{BSEL}
\begin{split}
    L(1,2,3,4)&= L_0(1,2,3,4)\\
    &+ L_0(1,5,8,4)I(5,6,7,8)L(6,2,3,7),
\end{split}
\end{equation}
where $L_0$ describes the independent propagation of a particle from point 1 to point 2 and a hole from point 4 to point 3,
\begin{equation}
L_0(1,2,3,4)= G(2,1)G(4,3),
\end{equation}and $I$ is the irreducible interaction kernel in the $GW$ approximation \cite{Hedin1965,Hanke1979,Strinati1988} 
\begin{equation}
\begin{split}\label{apprint}
I(1,2,3,4)=&\delta(1,4)\delta(2,3)v(1,2)\\
&-\delta(1,2)\delta(4,3)W(1,4).
\end{split}
\end{equation}
The interaction kernel consists of the repulsive bare Coulomb interaction $v(1,2)$ and the frequency-dependent screened interaction $W(1,2)$. In Eq.\ \eqref{BSEL} integration over repeated indices is implied. The first and second term of the irreducible interaction are obtained by varying the Hartree potential or the exchange and correlation part of the self-energy, respectively, with respect to the Green's function.

At this point, we make the static approximation to the screened interaction $W(1,2)$ \cite{Marini2003}. In this approximation, the full frequency-dependent screened interaction is approximated by its static value: $W(1,2)\approx W(\bm{r}_1,\bm{r}_2,\omega=0)$. Using this approximation the irreducible interaction $I$ is frequency-independent and equation \eqref{BSEL} can be solved in frequency space for $L(\omega)$:
\begin{equation}\label{Pminus1}
  L(\omega)^{-1}=L_0(\omega)^{-1} -I.  
\end{equation}
Until now all equations involved quantities which are continuous functions of space. To cast Eq.\ \eqref{Pminus1} into a matrix form, one needs to introduce a basis. Since we consider neutral electron-hole excitations, a suitable basis are the products of occupied and unoccupied orbitals, called resonant and antiresonant two-orbital states
\begin{equation}
\begin{split}\label{eq:two-orbital}
\Phi_K^{\text{r}}(\bm{r},\bm{r'}) &= \varphi_i (\bm{r}) \varphi^*_a(\bm{r'})\\
 \Phi_K^{\text{a}}(\bm{r},\bm{r'}) &= \varphi_a (\bm{r}) \varphi^*_i (\bm{r'}),
\end{split}
\end{equation}
where the indices $i,j,\dots$ and $a,b,\dots$ enumerate occupied and unoccupied states, respectively. Here and in the following we use the notation $K=\{i,a\}$ and $J=\{j,b\}$. In general, one also needs to include a $\bm k$-point index, $K=\{i\bm{k},a\bm{k}\}$, $J=\{j\bm{k}',b\bm{k}'\}$, but for brevity we will suppress it in the following. It can be easily restored, by adding $\bm{k}$ to $i$ and $a$, and $\bm{k}'$ to $j$ and $b$.

In this basis the resonant-resonant and antiresonant-antiresonant matrix elements of $L_0$ take the form
\begin{align}
(L_0(\omega)^{-1})^{(\text{r,r})}_{KJ}&= (\omega-(\epsilon_a-\epsilon_i))\delta_{ij}\delta_{ab}\label{L0resres},\\
(L_0(\omega)^{-1})^{(\text{a,a})}_{KJ}&= (-\omega-(\epsilon_a-\epsilon_i))\delta_{ij}\delta_{ab}\label{L0aresares}.
\end{align}
The resonant-resonant matrix elements of the irreducible interaction are then written
\begin{equation}\label{Amatrixelements}
\begin{split}
&\mathcal{H}_{KJ}^{(\text{r,r})}\!=\!\int\! d\bm{r}_1 \dots\, d\bm{r}_4 \\
&\times {\Phi_K^{\text{r}}}^*(\bm{r}_2,\bm{r}_4) I(\bm{r}_1,\bm{r}_2,\bm{r}_3,\bm{r}_4) \Phi_J^{\text{r}}(\bm{r}_3,\bm{r}_1),
\end{split}
\end{equation}
the matrix elements $\mathcal{H}_{KJ}^{(\text{r,a})}$ and $\mathcal{H}_{KJ}^{(\text{a,a})}$ are obtained analogously. Using the more compact Dirac notation we list here all anti-symmetrized two electron integrals corresponding to the interaction $I$:
\begin{align}
\mathcal{H}_{KJ}^{(\text{r,r})}&= \langle b i | V | j a \rangle-\langle b i | W | a j \rangle\label{Iresres} ,   \\
\mathcal{H}_{KJ}^{(\text{r,a})}&= \langle j a | V | b i \rangle-\langle j a | W | i b \rangle , \\
\mathcal{H}_{KJ}^{(\text{a,r})}&= \langle j i | V | b a \rangle - \langle j i | W | a b \rangle , \\
\mathcal{H}_{KJ}^{(\text{a,a})}&= \langle b a | V | j i \rangle -\langle b a | W | i j  \rangle\label{Iaresares} .
\end{align}
We have now arrived at a matrix representation of \eqref{Pminus1}:
\begin{equation}\label{eq:bsematrix}
\begin{split}
(L(\omega)^{-1})_{KJ} &= 
\omega
  \begin{pmatrix}
\mathbb{1} & 0 \\
0 & -\mathbb{1}\\
  \end{pmatrix} 
  -
  (\epsilon_a-\epsilon_i)
    \begin{pmatrix}
\delta_{ij}\delta_{ab} & 0 \\
0 & \delta_{ij}\delta_{ab} \\
  \end{pmatrix} \\
  &-
    \begin{pmatrix}
\mathcal{H}_{KJ}^{(\text{r,r})}&\mathcal{H}_{KJ}^{(\text{r,a})} \\
\mathcal{H}_{KJ}^{(\text{r,a})*}& \mathcal{H}_{KJ}^{(\text{r,r})*} \\
  \end{pmatrix}.
\end{split}
\end{equation}
Elementary excitations are determined by the poles of $L(\omega)$, in other words, those frequencies $\Omega$ for which the right hand side of Eq.\ \eqref{eq:bsematrix} is not invertible. This in turn means that the kernel of the matrix on the right hand side of Eq.\  \eqref{eq:bsematrix} is non-trivial, using the shorthand notation
\begin{equation}
\begin{split}
&
    \begin{pmatrix}
A & B \\
B^* &A^* \\
  \end{pmatrix}
= \\
&
  (\epsilon_a-\epsilon_i)
    \begin{pmatrix}
\delta_{ij}\delta_{ab} & 0 \\
0 & \delta_{ij}\delta_{ab} \\
  \end{pmatrix}
+  
    \begin{pmatrix}
\mathcal{H}_{KJ}^{(\text{r,r})}&\mathcal{H}_{KJ}^{(\text{r,a})} \\
\mathcal{H}_{KJ}^{(\text{r,a})*}& \mathcal{H}_{KJ}^{(\text{r,r})*} \\
  \end{pmatrix} ,
\end{split}
\end{equation}
this argument leads us to a generalized eigenvalue problem
\begin{equation}\label{generaleigen}
    \begin{pmatrix}
A & B \\
B^* &A^* \\
  \end{pmatrix}
        \begin{pmatrix}
X \\
Y\\
  \end{pmatrix}
  =
  \Omega
    \begin{pmatrix}
    \mathbb{1} & 0 \\
0 & -\mathbb{1}\\
  \end{pmatrix} 
        \begin{pmatrix}
X \\
Y\\
  \end{pmatrix}.
\end{equation}
For real-valued symmetric matrices $A$ and $B$, methods exist to solve this generalized eigenvalue problem \cite{Stratmann1998,Furche2001}. For complex-valued matrices, one can take advantage of the time inversion symmetry of Bloch states and reorder the antiresonant states in the second row and column to transform $B^*$ and $A^*$ to $B$ and $A$, respectively \cite{Sander2015}. This way one can reduce the 2N non-Hermitian eigenvalue problem into two diagonalizations of Hermitian matrices of size N, where N is the number of particle-hole pairs \cite{Sander2015}.

However, in the present work we use the well-known \emph{Tamm-Dancoff} approximation \cite{Dancoff1950,Tamm1991}, i.e.\ neglect the off-diagonal matrices $B,B^*$. We further use the analytical property that response functions derived from $L(\omega)$ are even functions of $\omega$ and the simplification that for Bloch states $A=A^*$ \cite{Sander2015}. All of this reduces the preceding generalized eigenvalue problem in Eq.\ \eqref{generaleigen} to a simple eigenvalue equation for a Hermitian matrix $A$
\begin{equation}\label{BSEEV}
A X = \Omega X.
\end{equation}

We have now summarized the BSE formalism and its main ingredients: the matrix elements of the resonant-resonant matrix $A$, equation \eqref{Amatrixelements} and the eigenvalue equation \eqref{BSEEV}. In the next section we summarize the pertinent parts of the PAW method: the basic principles and how core states are treated. 

\subsection{Basics of the PAW formalism}\label{sec:PAW}

The PAW method is an all-electron (AE) method in which the exact orbital $|\psi\rangle $ is obtained from the pseudo (PS) orbital $|\tilde \psi\rangle$ via a linear transformation \cite{Bloechl1994}
\begin{equation}\label{PAW_transform}
|\psi\rangle = |\tilde \psi\rangle + \sum_n  c_n | \phi_n^{1} \rangle -  c_n  |\tilde \phi_n^{1}\rangle,
\end{equation}
where $|\tilde \psi\rangle$ are the PS orbitals represented on a plane-wave grid, these are the variational quantities. $|\phi^{1}\rangle$ and $|\tilde\phi^{1}\rangle$ are all-electron partial waves and PS partial waves, respectively, both defined on a radial grid. We use a superscript $1$ to denote one-center quantities located inside PAW spheres and evaluated on a radial grid. The coefficients $c_n$ are projections of the pseudo orbitals on projectors defined inside the PAW spheres
\begin{equation}
c_n=\langle\tilde p_n | \tilde\psi\rangle.
\end{equation}
The index $n$ is a shorthand for an atomic site index $\tau_n$, angular $l_n$ and magnetic quantum numbers $m_n$, as well as an additional index for the reference energy $\epsilon_n$. Here and in the following, compound indices $n$ and $m$ are used to index the projectors, partial waves, and coefficients $c_n$. 
The on-site expansion $\sum_n c_n  |\tilde \phi_n^{1}\rangle$ must be equal to $|\tilde \psi\rangle$ inside the PAW spheres, which implies that 
\begin{equation}\label{complete}
\sum_n |\tilde\phi_n^1\rangle\langle \tilde p_n |=1.
\end{equation}
For core states, the coefficients in the  transformation \eqref{PAW_transform} are unity \cite{Bloechl1994}
\begin{equation}\label{PAW_core}
|\psi_c\rangle =| \tilde\psi_c\rangle+ | \phi_c ^1\rangle -| \tilde\phi_c ^1\rangle.
\end{equation}
In practice, one can safely assume that all quantities in the above equation are entirely localized inside the PAW spheres, hence $| \tilde\psi_c\rangle = |\tilde \phi_c ^1\rangle$ and $|\psi_c\rangle = | \phi_c ^1\rangle $. Furthermore, we will make the assumption that $| \tilde\psi_c\rangle = |\tilde \phi_c ^1\rangle =0$. This is justified, since the core orbitals do not contribute outside of the PAW spheres, and in the PAW method one is free to make any choice for the pseudo partial waves $|\tilde \phi_c ^1\rangle$, as long as they are identical to the all-electron partial waves $|\phi_c ^1\rangle$ outside the PAW spheres. Since $|\phi_c ^1\rangle$ is zero outside of the PAW spheres, and hence $|\tilde \phi_c ^1\rangle$ is also zero outside the PAW spheres, one can also assume that $\tilde \phi^1_c(\bm{r})=0$ and thus $\tilde \psi_c(\bm{r})=0$ everywhere in space. We will give further support to this argument  towards the end of the next subsection.

Inside the PAW spheres the AE partial waves  are  solutions of the radial Schrödinger equation for a specific energy $\varepsilon_n$ and angular momentum quantum numbers $l_n,m_n$
\begin{equation}\label{AE_partial}
  \phi_n^1(\bm{r})=\frac{1}{r}u_{\epsilon_n,l_n}(r)Y_{l_n,m_n}(\theta,\phi),   
\end{equation}
with radial functions $u(r)$ and   spherical harmonics $Y_{l,m}(\theta,\phi)$. These are calculated for isolated atoms when the PAW potentials are generated.  The AE core orbitals are also calculated for the Kohn-Sham potential of the isolated atom, imposing the boundary condition that core orbitals become zero at the radius of the PAW sphere \cite{Bloechl1994,Kresse1999}. This is done in a preprocessing step within the VASP code.
Note that except for a Bloch phase-factor $e^{i \bm{k}\cdot\bm{r}}$ the core orbitals are identical at all $\bm k$-points .

The PS partial waves $\tilde{\phi}^1_n(\bm{r})=\langle \bm{r}|\tilde{\phi}_n^1\rangle$ are analogously given as
\begin{equation}\label{PS_partial}
  \tilde \phi_n^1(\bm{r})=\frac{1}{r}\tilde u_{\epsilon_n,l_n}(r)Y_{l_n,m_n}(\theta,\phi),
\end{equation}
and determined by pseudizing the AE partial waves inside a suitably chosen core radius. 

In the next section, we describe in detail how to evaluate the matrix elements in the PAW method, and how the expressions need to be modified when core states are included in the transitions.

\subsection{BSE matrix elements within PAW}\label{sec:matelements}

We now evaluate the two electron integrals in Eq.\ \eqref{Amatrixelements}. Defining charge densities 
\begin{equation}
 n_{ab}(\bm{r})=\varphi_a^* (\bm{r}) \varphi_b(\bm{r}),
 \end{equation}
the two terms of the resonant-resonant matrix elements, Eq.\ \eqref{Iresres}, can be written as
\begin{align}
\langle b i | V | j a \rangle&= \int d\bm{r} d\bm{r}' n_{ia}(\bm{r}) v(\bm{r}-\bm{r}')n_{jb}^*(\bm{r}'),\label{eq:exc_matrixelements}\\
 \langle b i | W | a j \rangle&= \int d\bm{r} d\bm{r}' n_{ij} (\bm{r}) W(\bm{r},\bm{r}') n_{ab}^*(\bm{r}').\label{eq:dir_matrixelements}
\end{align}
Following convention, we refer to the terms involving $v$ and $W$ as exchange and direct terms, respectively \cite{Onida2002}.

To calculate the matrix elements of Eqs.\ \eqref{eq:dir_matrixelements} and \eqref{eq:exc_matrixelements} we resort to the formalism of augmentation charges \cite{Bloechl1994,Kresse1999}. 
Augmentation charges are constructed such that inside the PAW spheres the sum of PS charge density and the augmentation charge $\hat n$ has the same moments as the exact charge density
\begin{equation}
\label{equ:multipole}
    \int_{\Omega_r} [n^1(\bm{r}) - \tilde n^1(\bm{r}) - \hat n(\bm{r})] |\bm{r}|^l Y^*_{lm}(\theta,\phi)  d\bm{r}=0,
\end{equation}
where the coordinate system is centered on a particular PAW sphere. Details on the explicit construction of the augmentation charges can be found in Ref.\ \cite{Kresse1999}.
With this definition of the augmentation charge, we write the exact charge density as a sum of three terms:
\begin{equation}\label{charge_decomposition}
\begin{split}
            n_{ab}(\bm{r}) & = [\tilde n_{ab}(\bm{r}) + \hat{n}_ {ab}(\bm{r})]  \\
                 &     + n^1_{ab}(\bm{r})  - [\tilde n^1_{ab}(\bm{r}) + \hat{n}^1_{ab}(\bm{r})],
\end{split}
\end{equation}
where the first term is the plane-wave charge density plus the augmentation charge on the regular grid, the second term is the AE charge density, and the last term is the PS charge density plus PS augmentation charge on the radial grid.

At this point we make an important approximation and neglect all one-center terms, i.e., the second line of \eqref{charge_decomposition}:
\begin{equation}\label{charge_approx}
    n_{ab}(\bm{r})\approx [\tilde n_{ab}(\bm{r}) + \hat{n}_ {ab}(\bm{r})].
\end{equation}
This approximation is made consistently in the VASP code for the $GW$, RPA, and BSE implementations.
Using this approximation, we can write an explicit expression for the exchange and direct terms, respectively:
\begin{equation}\label{eq:dir_exc_matrixelements_augment1}
\begin{split}
    \langle b i | V | j a \rangle&= \int d\bm{r} d\bm{r}' [\tilde n_{ia}(\bm{r}) + \hat{n}_ {ia}(\bm{r})] \\
    &\times v(\bm{r}-\bm{r}')[\tilde n_{jb}(\bm{r}') + \hat{n}_ {jb}(\bm{r}')]^*,
\end{split}
\end{equation}
\begin{equation}\label{eq:dir_exc_matrixelements_augment2}
\begin{split} 
\langle b i | W | a j \rangle &= \int d\bm{r} d\bm{r}' [\tilde n_{ij}(\bm{r}) + \hat{n}_ {ij}(\bm{r})]\\
&\times W(\bm{r},\bm{r}') [\tilde n_{ab}(\bm{r}') + \hat{n}_ {ab}(\bm{r}')]^*.
\end{split}
\end{equation}
These expressions have already been implemented previously for the case of transitions from valence to conduction band states \cite{Sander2015}.

We now discuss the case of core states in some more detail.
If one of the occupied states $i$ above is a core state $c$ then there are three combinations of orbitals possible: $n_{cc'}$ (two core states), $n_{ci}$ (a core state and a valence band state), or $n_{ca}$ (a core state and a conduction band state). For a core state $c$ inside a particular PAW sphere, this core state is confined entirely to the sphere and vanishes beyond this PAW sphere. It is then evident that each of the three charge densities mentioned above contributes only inside the PAW sphere in which the core state $c$ is located. 

We show here how this implies that the plane-wave contributions $\tilde{n}$ in Eqs.\ \eqref{eq:dir_exc_matrixelements_augment1} and \eqref{eq:dir_exc_matrixelements_augment2} vanish.
We first use the formula for the expectation value of a local operator in the PAW scheme for the real-space projection operator $|\bm{r}\rangle\langle \bm{r}|$, see Eq.\ (11)  of Ref.\ \cite{Bloechl1994}:
\begin{equation}\label{eq:pawchargedensity}
\begin{split}
        \langle \varphi_i| \bm{r}\rangle \langle \bm{r} |\varphi_b\rangle &= \langle\tilde\varphi_i | \bm{r} \rangle \langle  \bm{r} |\tilde\varphi_b  \rangle \\
        & +\sum_{nm} \langle\tilde\varphi_i | \tilde p_n \rangle \langle \tilde p_m |\tilde\varphi_b \rangle \langle \phi_n^1 | \bm{r}\rangle \langle \bm{r} |\phi_m^1\rangle \\
        & -\sum_{nm} \langle\tilde\varphi_i | \tilde p_n \rangle \langle \tilde p_m |\tilde\varphi_b \rangle \langle \tilde\phi_n^1 | \bm{r}\rangle \langle \bm{r} |\tilde\phi_m^1\rangle.
\end{split}
\end{equation}
We now restrict the state $i$ to a core state $c$, while the state $b$ is a conduction band state. As argued above, the charge density $\langle \varphi_c| \bm{r}\rangle \langle \bm{r} |\varphi_b\rangle$ has contributions inside the PAW spheres only. Hence, we can use the completeness relation in Eq.\ \eqref{complete}, and the first and third terms of Eq.\  \eqref{eq:pawchargedensity} cancel. This finally implies that the plane wave charge densities $\tilde n$ in Eqs.\ \eqref{eq:dir_exc_matrixelements_augment1} and \eqref{eq:dir_exc_matrixelements_augment2} can be neglected, so that the exchange and direct terms can be written as
\begin{equation}\label{eq:dir_exc_matrixelements_augment}
\langle b c | V | c a \rangle = \int d\bm{r} d\bm{r}' \hat{n}_ {ca}(\bm{r}) v(\bm{r}-\bm{r}')\hat{n}^*_ {cb}(\bm{r}')
\end{equation}
and
\begin{equation}
\begin{split}
 \langle b c | W | a j \rangle&= \int d\bm{r} d\bm{r}' \hat{n}_ {cj}(\bm{r}) \\
 &\times W(\bm{r},\bm{r}') [\tilde n_{ab}(\bm{r}') + \hat{n}_ {ab}(\bm{r}')]^*,
 \end{split}
\end{equation}
respectively.
As shown in Eq.\ \eqref{equ:multipole}, the augmentation charges are constructed in such a way that they restore the exact multipoles of the all-electron charge density. This implies that even though we use only augmentation charge densities, the long range electrostatic effects are exactly accounted for. The argument laid out here also confirms that  $\tilde \phi^1_c(\bm{r})= \tilde \psi_c(\bm{r})=0$ everywhere in space is a valid choice. The expressions above have been implemented in the \textsc{VASP} code for core-conduction as well as core-valence transitions. 

\subsection{Transition probabilities and dielectric function}\label{methods_dielectric}

The dielectric function is finally calculated as 
\begin{equation}\label{eq:DF}
\begin{split}
  \varepsilon_{\text{M}}(\bm{q},\omega) & = 1 + \lim_{\bm{q} \to 0}\, v(\bm{q})
 \sum_{\Lambda} \left(\frac{1}{\Omega_{\Lambda} -\omega } + \frac{1}{\Omega_{\Lambda}+ \omega } \right)\times\\
&   1/{N_k} \Bigg\lbrace \sum_{\bm k} \sum_{a,c} 
\langle a{\bm k}|e^{i\bm{q}\cdot \bm{r}}|c{\bm k} \rangle  
 X_{\Lambda}^{(c{\bm k},a{\bm k})} \Bigg\rbrace \Bigg\lbrace c.c.\Bigg\rbrace,
\end{split}
\end{equation}
compare Eqs.\ (50) and (51) of Ref.\ \cite{Sander2015}. Here, $N_k$ is the total number of $\bm{k}$-points, and 
the index $\Lambda$ labels the eigenstates of the eigenvalue problem in Eq.\ \eqref{BSEEV}. $v(\bm{q})$ is the Coulomb kernel in CGS units, $v(\bm{q})= (1/V) 4 \pi e^2/ |\bm{q}|^2$, where $V$ is the volume of the unit cell. For clarity, the $\bm{k}$-point index was added back. Although the present implementation allows for a simultaneous treatment of core-conduction and valence-conduction pairs, we did not consider the valence states in our present BSE calculations. Then, the key ingredient here is the transition probability between core and valence-band states:
\[
 \lim_{\bm{q} \to 0} \sqrt{v(\bm{q})} \langle a{\bm k}|e^{i\bm{q}\cdot \bm{r}}|c{\bm k} \rangle,
\] 
which in  $\bm{k}\cdot\bm{p}$ perturbation theory is approximated as
\[
   \sqrt{4 \pi e^2/V} \langle a{\bm k}|  \nabla |c{\bm k} \rangle / ( \epsilon_{c } - \epsilon_{a{\bm k}}).
\]
The $\nabla$-term is evaluated inside the PAW sphere between the exact all-electron partial waves of the core orbitals  
and the all-electron partial waves corresponding to the considered conduction band state $c{\bm k}$.
The required core eigenvalues $\epsilon_{c}$ are assumed to be identical to the DFT core eigenvalues and identical for all $\bm k$-points,  whereas the conduction band energies $\epsilon_{a{\bm k}}$ are the approximate QP energies in the GW calculations.
VASP routinely calculates the DFT core eigenvalues by evaluating the expectation value of the frozen core
orbitals in the self-consistent Kohn-Sham potential. Alternatively, one can specify the core eigenvalue as input.
However, results are largely independent of $\epsilon_{c}$.
Changing $\epsilon_{c}$ changes the onset of absorption and scales all intensities by roughly a constant value. Since the onset of  absorption is not yet accurately predicted by DFT or even $GW$, we use this freedom to adjust the onset of the absorption spectrum to the experimental value.

\section{Computational Methods}\label{sec:methods}
\subsection{Numerical details}\label{methods_general}
All \emph{ab initio} electronic structure calculations were performed with the all-electron plane-wave  code VASP \cite{Kresse1996}, which uses the PAW-implementation of Kresse and Joubert \cite{Kresse1999}. The workflow is the following.

We start with a standard DFT calculation, yielding the Kohn-Sham energies and Kohn-Sham orbitals. In all DFT calculations we used the exchange-correlation functional by Perdew, Burke, and Ernzerhof \cite{Perdew1996}. To calculate the quasiparticle energies and the dynamic screened interaction in momentum space $W_{\bm{G},\bm{G}'}(\omega)$, we perform a single-shot $G W$ calculation ($G_0 W_0$). The static approximation for $W$ is made at the beginning of the BSE calculation. Note that the orbitals are kept fixed at the DFT level. In all systems considered, the energy cutoff of the response function was set to $150$ eV. Then, the BSE eigenvalue equation is set up using the $GW$ quasiparticle energies and the PBE orbitals. Finally, the BSE equation is solved and the BSE dielectric function is calculated.

We reiterate that in this work we only investigate $K$-edges.  Furthermore, transitions from valence states to conduction states are excluded from the calculations. The number of conduction bands included in the transitions is $8$ in diamond, $12$ in graphite, $12$ in $h$-BN, and $15$ in all four lithium-halides. 

To reduce the computational demand and still obtain highly accurate spectra, we employ the shifted grid technique in all $GW$+BSE calculations  \cite{Sander2015}. First, we generate all irreducible $\bm{k}$-points $\bm{k}_{1,\dots,L}$ and corresponding weights $w_{1,\dots,L}$ of an $n\times n\times n$ $\bm{k}$-mesh. Then, we perform $L$ independent calculations with an $m\times m\times m$ $\bm{k}$-mesh, where the k-point grid is shifted by one of the $L$ irreducible $\bm{k}$-points. This procedure generates all $\bm{k}$-points of a regular ${(m\cdot n)} \times (m\cdot n)  \times (m\cdot n) $ $\bm{k}$-mesh. Finally, we average the $L$ so obtained dielectric functions, $\varepsilon={\sum_i^L w_i\varepsilon_i}/{\sum_i^L w_i}$.

For diamond, graphite, and $h$-BN we compare the $GW$+BSE spectra to those of the SCH calculations implemented previously \cite{Karsai2018}. In all SCH calculations the supercells contain $128$ atoms. In Table \ref{kpointstable}, we list the $\bm{k}$-meshes used in the $GW$+BSE and SCH calculations.

\begin{table}[]
    \centering
\begin{tabular}{l c c l c c}
\hline\hline
System & BSE & BSE & BSE & SCH  & SCH \\
System & $n$ & $L$ &$m\times m\times m$ & atoms  & $\bm{k}$-points \\
\hline
diamond &  $3$ & $4$ & $10\times 10\times 10$  & 128 &9$\times$9$\times$9\\
graphite &  $4$ & $12$ & $16\times 16\times 4$ & 128 &4$\times$4$\times$2\\
$h$-BN &  $3$ & $4$ & $12\times 12\times 4 $   & 128 &5$\times$5$\times$3\\
lithium-halides &  $3$ & $4$ & $10\times 10\times 10$& 128 & 8$\times$8$\times$8\\
\hline\hline
\end{tabular}
    \caption{Table of $\bm{k}$-meshes used for the shifted grid technique and for the supercell core-hole calculations (SCH). Shifted grid technique: an $n\times n \times n$ mesh results in $L$ irreducible $\bm{k}$-points. $m\times m\times m$ meshes are then shifted by these $L$ different $\bm{k}$-points, creating an $(m\cdot n) \times (m\cdot n)  \times (m\cdot n) $ $\bm{k}$-mesh. All meshes are centered on the $\Gamma$-point. For the SCH calculations we specify both the supercell size (number of atoms), as well as the $\bm{k}$-point grids.}
    \label{kpointstable}
\end{table}

In the available experimental absorption spectra of graphite \cite{Brandes2008} and of nitrogen in $h$-BN \cite{Li2012}, the incoming radiation was incident at an angle $\alpha$ of $40^\circ$ and $45^\circ$ to the surface normal, respectively. In these cases, we have mixed perpendicular and parallel components of the dielectric function according to $\text{Im}(\varepsilon) \sim \cos(\alpha)^2 \varepsilon_{xx}+\sin(\alpha)^2 \varepsilon_{zz}$, where $\varepsilon_{xx},\varepsilon_{yy}$ are the in-plane and $\varepsilon_{zz}$ the  out-of-plane components of the dielectric tensor. Details on the polarization are missing for the $K$-edge of nitrogen in $h$-BN \cite{Petravic2013}, for this system we have weighted each component by a factor 1/3.

In all the presented spectra for C, B, and N, an energy-independent Lorentzian broadening of 0.3 eV is applied. In the lithium-halides, a reduced broadening of 0.1 eV is used.

Finally, we note that neither the $GW$+BSE nor the SCH method can quantitatively calculate excitation energies, hence the energy axes of the modeled spectra are always shifted in order to obtain best agreement with the experimental spectra. 

\subsection{Model dielectric function approach}

In the $GW$+BSE approach, the screened interaction $W$ is the output of the $G_0W_0$ step. Since we perform single-shot $GW$ calculations, $W$ is calculated in the RPA approximation from the PBE orbitals and eigenvalues. 
$GW$ calculations are computationally demanding, scaling with the fourth power of the system size and quadratic with the number of $\bm{k}$-points \cite{Liu2016}. We investigate whether $W$ can be obtained cheaper and still sufficiently accurate by using the model dielectric function approach \cite{Bokdam2016}. In this approach, a standard DFT calculation is done as a preparatory step. Then, the $G_0W_0$ step is skipped and the screened interaction is instead calculated via $W=\varepsilon^{-1}v$, where the diagonal inverse dielectric function $\varepsilon^{-1}$ is approximated by the model dielectric function 
\begin{equation}\label{modeldielectric}
    \varepsilon_G^{-1}(\varepsilon_\infty^{-1},\mu)=1-(1-\varepsilon_\infty^{-1})e^{-G^2/4\mu^2},
\end{equation}
with $G=|\bm{G}|$. Here, the macroscopic dielectric function  $\varepsilon_\infty$ is obtained by averaging the diagonal elements of the macroscopic dielectric tensor, the screening parameter $\mu$ results from a fit of the model-dielectric function to $\varepsilon_{G}^{-1}(\bm{q}\to 0, \omega=0)$, where the long-range and short-range limits are set to $1/\varepsilon_\infty$ and 1, respectively.

This approach is used for diamond and B in $h$-BN in addition to the $GW$+BSE spectra. Furthermore, we need to assess how well this approach approximates the diagonal elements of $W$. To do this, we also show for these two systems $GW$+BSE spectra where the off-diagonal elements of $W$ have been set to zero in the BSE step. 

In summary, we compare for diamond and B in $h$-BN three spectra: a $GW$+BSE spectrum using the full screened interaction including off-diagonal elements, a $GW$+BSE spectrum using only the diagonal elements of $W$, and finally a BSE spectrum using the model dielectric function approach.

\section{Results}\label{sec:results}

\subsection{Diamond}

\begin{figure}
\centering
\includegraphics[scale=1]{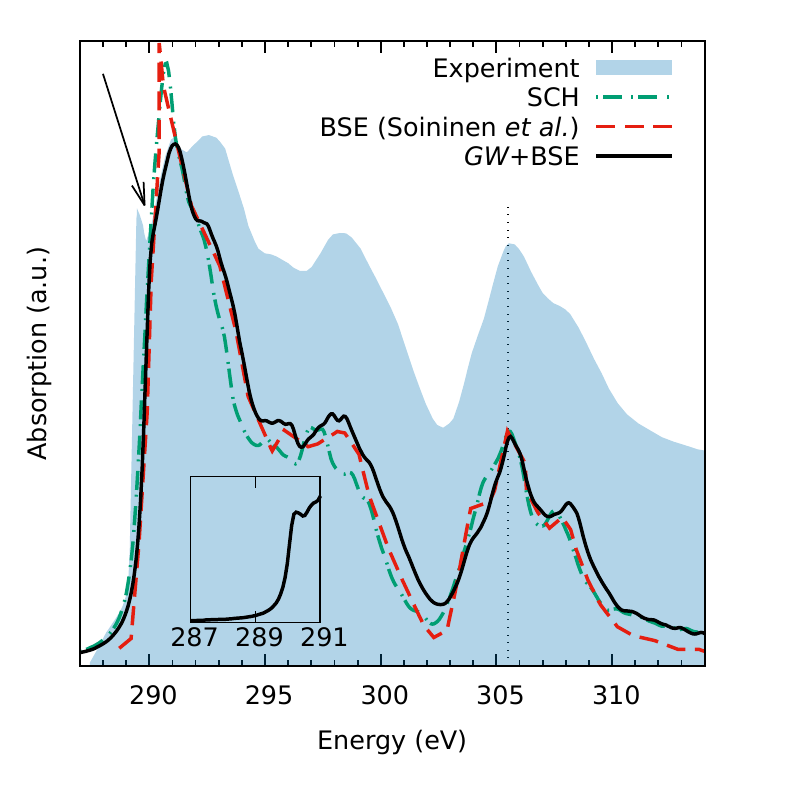}
\caption{XAS spectra of $K$-edge of $C$ in diamond. \emph{Blue filled curve:} experiment \cite{Ma1992}. \emph{Black curve:} $GW$+BSE spectrum, this work. \emph{Green dash-dotted curve:} SCH calculation. \emph{Red dashed line:} BSE-spectrum from \cite{Soininen2001}. \emph{Arrow:} Location of the excitonic shoulder in the $GW$+BSE spectrum. \emph{Inset:} $GW$+BSE result using an $18\times 18 \times 18$ $\bm{k}$-mesh and a reduced Lorentzian broadening. All modeled XAS-spectra are centered and adjusted in scale on the peak at at $305.5$ eV. 
}
\label{diamond3}
\end{figure}

\begin{figure}
\centering
\includegraphics[scale=1]{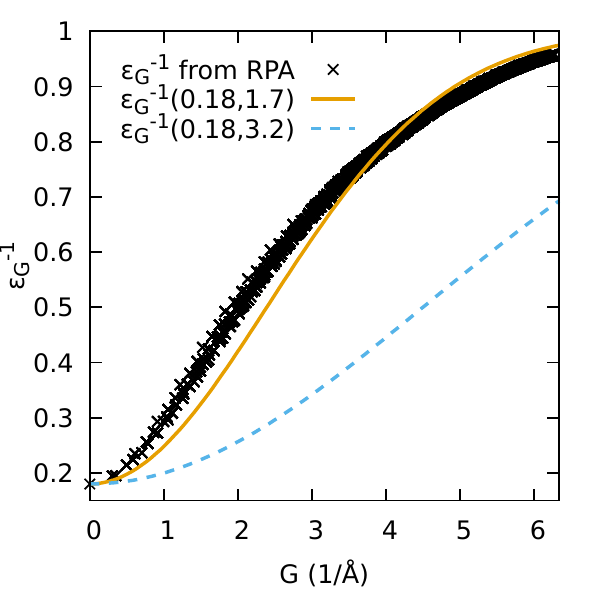}
\caption{Plot of the inverse dielectric function $\varepsilon_G^{-1}$ of diamond. \emph{Crosses:} Inverse dielectric function $\varepsilon_G^{-1}(\bm{q}\to 0, \omega=0)$ at the RPA level. \emph{Orange line:} model dielectric function $\varepsilon_G^{-1}(0.18, 1.7)$. \emph{Blue dashed line:} $\varepsilon_G^{-1}(0.18, 3.2)$. All lines intercept the $y$-axis at $\varepsilon_\infty^{-1}=0.18$.}
\label{diamond2}
\end{figure}

\begin{figure}
\centering
\includegraphics[scale=0.9]{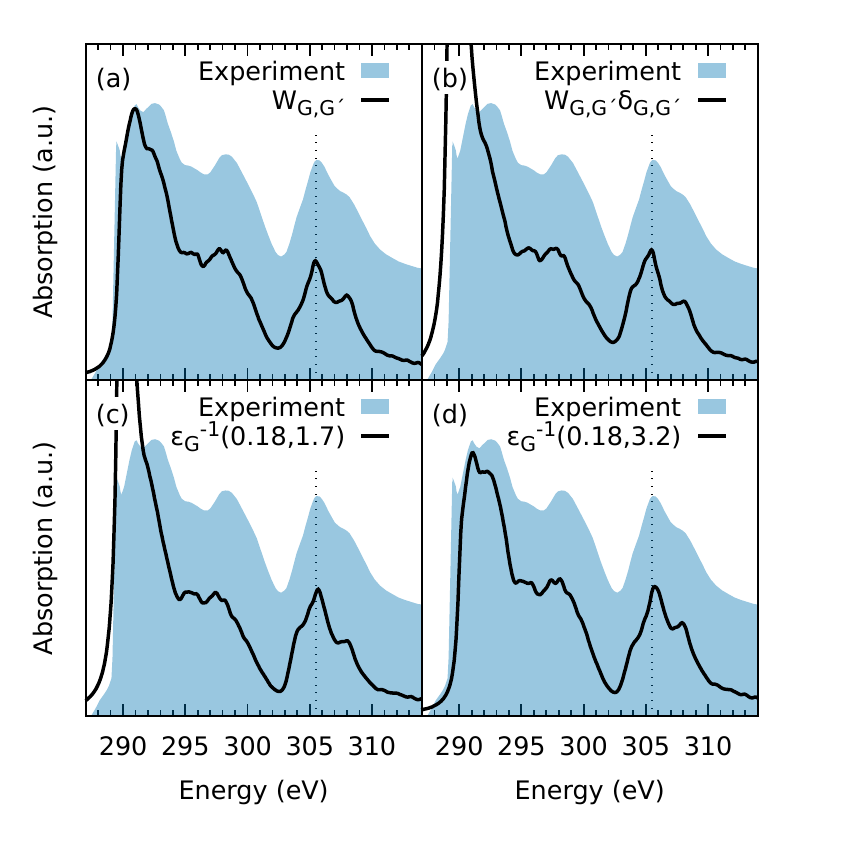}
\caption{$K$-edge spectrum of $C$ in diamond, comparing experiment \cite{Ma1992} (\emph{blue filled curve}) to $GW$+BSE spectra, using different approaches to the screened interaction $W$. (a):  $W$ from $G_0 W_0$ step. (b):  same as panel (a), excluding off-diagonal elements. (c): Model dielectric function $\varepsilon_G^{-1}(0.18, 1.7)$. (d): Model dielectric function $\varepsilon_G^{-1}(0.18, 3.2)$, $\mu$ adjusted manually.}
\label{diamond1}
\end{figure}

In Fig.\ \ref{diamond3} we compare the $K$-edge of diamond using the present PAW--$GW$+BSE implementation with experiment \cite{Ma1992}, a pseudopotential BSE implementation \cite{Soininen2001}, and an SCH-spectrum. Comparing the modeled spectra to experiment we see that the positions of the peaks agree very well, but the spectrum lacks intensity towards higher energies. Interestingly, the difference between the theoretical and experimental spectra seems to be mostly a "background" present in the experimental spectrum: for instance the intensity modulations around the individual peaks follow the experimental intensity modulations exceedingly well. This is most obvious around the minimum at 302 eV: both to the left and right hand side, the experimental and theoretical spectrum change by roughly the same amount. The background consistently increases from the leftmost first peak towards the right, and can be roughly modelled by a parabolic curve with the onset of the parabola located around the first peak. The same discrepancy will be noticeable for all spectra that we show in the present work. We are not certain about the origin of the discrepancy, but speculate that it is related to multiple scattering events involving core-conduction band pairs at other sites, or alternatively high energy valence-conduction band pairs. Both are  not  accounted for by the present level of theory.

Returning to the diamond spectrum, we note that in the experimental spectrum at around 290 eV a slight, most likely excitonic peak is visible. Compared to the other spectra our result shows a slight shoulder located at around the same energy (see arrow in Fig.\ \ref{diamond3}). To show that this feature is not an artifact of the shifted grid technique, we plot in an inset up to the first maximum at 291 eV the result of a calculation using a standard non-shifted $18\times18\times18$ $\bm{k}$-point mesh. Clearly, this spectrum also shows a peak. To make it better visible we have reduced the Lorenzian broadening in this calculation to 0.2 eV.

In summary, for the case of diamond the present spectrum agrees exceedingly well for all peak positions with the experimental spectrum. Every single peak of the experimental spectrum is resolved, even small ones, and deviations are only observed for the absolute intensities of the peaks. We have to keep in mind though that the experiments usually show some "background intensity", in particular above the onset of core excitations. This leads to a stronger absorption intensity further from the edge of the spectrum. 

In Fig.\ \ref{diamond1} we compare the experiment to four $GW$+BSE spectra, where in each panel we take a different approach to the screened interaction, as outlined in the computational methods section. All spectra are centered on and adjusted in scale to the peak of the experimental spectrum at $305.5$ eV, marked by a vertical dashed line.  Comparing the results obtained by including (panel (a)) or excluding (panel (b)) off-diagonal elements of $W$, we see that the magnitude of the first peak is vastly overestimated when only the diagonal elements are included. In panel (c) we use a model-dielectric function $\varepsilon_G^{-1}(0.18,1.7)$. As in panel (b), the first feature is overestimated. In panel (d) we use a model-dielectric function with unchanged $\varepsilon_\infty^{-1}$ and a screening parameter manually set to $\mu=3.2~\text{\AA}^{-1}$. Increasing $\mu$ restores the amplitude of the first peak to that of panel (a). 

Neglecting the off-diagonal elements of the screened interaction from panel (a) to panel (b) in Fig.\ \ref{diamond1} or only using the diagonal model-dielectric function in panel (c) overestimates the first peak and redshifts the peak at 297 eV slightly to the left. In Fig.\ \ref{diamond2} we plot the diagonal elements of the RPA dielectric matrix as well as the model dielectric functions $\varepsilon_G^{-1}(0.18,1.7)$ and $\varepsilon_G^{-1}(0.18,3.2)$. From this plot we see that increasing $\mu$  decreases, for a given wave vector $G$, $\varepsilon_G^{-1}$ and hence decreases $W=\varepsilon^{-1}v$. Or in short, increasing $\mu$ increases the screening at larger $G$ vectors.  By increasing the screening parameter $\mu$ from panel (c) to (d), the magnitude of the first peak is readjusted to the full results (panel (a)). 

From all these observations we conclude that the off-diagonal elements of the screened interaction are very relevant for an accurate description of excitonic peaks of localized core states.  Although one can mimic this effect by adjusting by hand the screening length in the model dielectric function, this remains a rather empirical approach, and clearly the model dielectric function after adjustment of $\mu$ does not follow the \emph{ab initio} results, compare Fig.\ \ref{diamond2}. As to why the off-diagonal components are relevant, we note that we made similar observations for small molecules and in general for strongly localized states. The diagonal approximation in momentum space is simply not adequate when one deals with a mixture of localized and itinerant states, whereas it accounts well for transitions involving only band-like itinerant states. This is an important observation that most likely exposes and underlines the limits and problems of a simple diagonal screening approach.

\subsection{Graphite}

\begin{figure}
\centering
\includegraphics[scale=0.9]{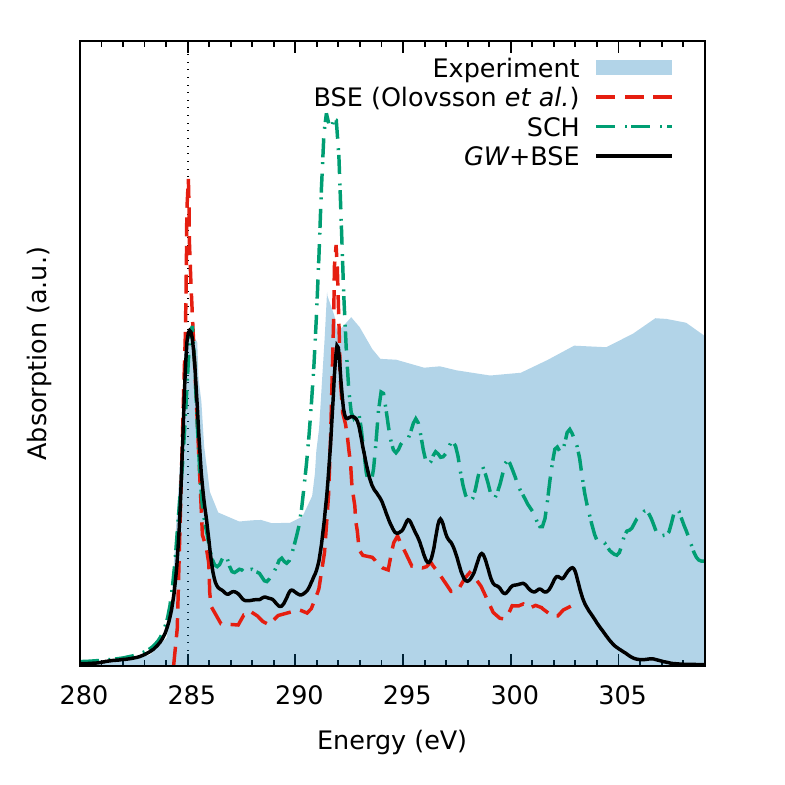}
\caption{XAS spectra of the $K$-edge of $C$ in graphite.    \emph{Blue filled curve:} experiment  \cite{Brandes2008}. \emph{Black curve:} $GW$+BSE spectrum, this work. \emph{Green dashed curve:} BSE-spectrum from \cite{Shirley2000}. \emph{Red dash-dotted curve:} SCH calculation.  All modeled spectra are centered at $285$ eV and adjusted in height to match the integral over the first peak. }
\label{graphite1}
\end{figure}

\begin{figure}
\centering
\includegraphics[scale=1]{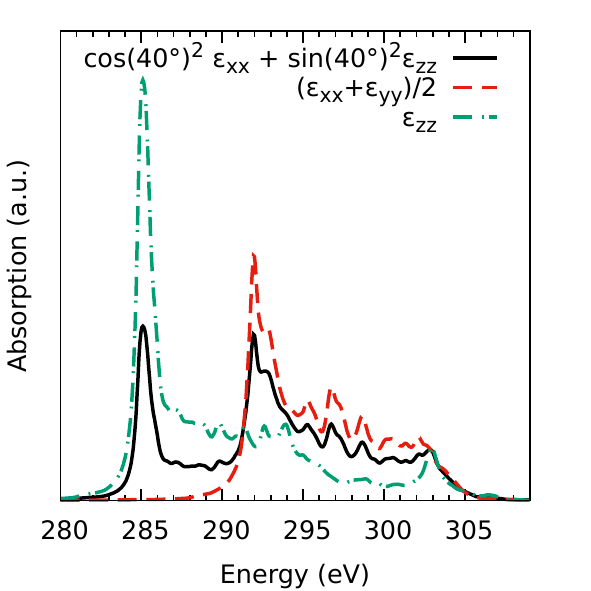}
\caption{Components of the dielectric tensor for the $K$-edge of $C$ in graphite. \emph{Black curve:} components averaged according to $\text{Im}(\varepsilon) \sim \cos(40^\circ)^2 \varepsilon_{xx}+\sin(40^\circ)^2 \varepsilon_{zz}$. \emph{Orange dashed curve:} in-plane components, $\frac{1}{2}(\varepsilon_{xx}+\varepsilon_{yy})$. \emph{Green dash-dotted curve:} out-of-plane component, $\varepsilon_{zz}$.
}
\label{graphite2}
\end{figure}

In Fig.\ \ref{graphite1} we present a comparison of our PAW--$GW$+BSE result of the $K$-edge of graphite to experiment \cite{Brandes2008}, to an all-electron full-potential BSE spectrum  \cite{Olovsson2019}, and an SCH-spectrum. In Ref.\ \cite{Olovsson2019} perpendicular and parallel components are shown separately, we have mixed them in the same way as our result. The $GW$+BSE and SCH peaks have been adjusted in height to the first peak in experiment. Since the spectrum of Olovsson \emph{et al.}\ \cite{Olovsson2019} has been calculated using less broadening, we have adjusted its height to approximately match the integral over the first peak. Comparing the PAW--$GW$+BSE spectrum of this work and the all-electron result of Ref.\ \cite{Olovsson2019} we see that both results can match the energy difference of the experimental peak positions at 285 eV and 292 eV.

Graphite is a 2D van-der-Waals material and we show the in-plane and out-of-plane components of the dielectric tensor in Fig.\ \ref{graphite2}: The peak at $285$ eV can be fully attributed to the out-of-plane component $\varepsilon_{zz}$. 

While our result as well as the spectrum from Ref.\ \cite{Olovsson2019} can reproduce the intensity ratio of the peaks at 287 eV and 292 eV reasonably well, the SCH spectrum overestimates the second peak. We also note the difference in amplitude of our result and experiment in the ranges 285-292 eV and from 292 eV upwards. Above 305 eV, we find no intensity, whereas the experimental signal remains quite substantial. As before, the experimental data show a substantial background at higher energies that is lacking in our theoretical calculations. We note that as for diamond the fine-structure of the experimental XAS spectrum is exceedingly well resolved in our theoretical calculations: note the peaks at 292, 293, 295, 296, and 303 eV are also visible in the experimental spectrum at least as slight humps.

It was suggested in Ref.\ \cite{Olovsson2019} that the double-peak structure at 292 eV and 293 eV is due to a breaking of degeneracy induced by the electron-phonon interaction. While we cannot exclude that electron-phonon coupling will enhance the splitting, we clearly observe two peaks even in the absence of any lattice distortions. In the previous BSE calculations of Ref.\ \cite{Olovsson2019}, this feature was not observed, most likely because of an insufficient $\bm{k}$-point sampling density. In our case, this feature is robust with respect to different $\bm{k}$-point samplings. In Ref.\ \cite{Olovsson2019}, the controversy around the $\sigma_2^*$ peak is discussed and it is mentioned that some studies interpret this peak as a delocalized bandlike contribution \cite{Bruehwiler1995}. If we interpret the shoulder in our spectrum tentatively as the $\sigma_2^*$ peak then an explanation might be given why we can see the shoulder in our spectrum and not in the work of Ref.\ \cite{Olovsson2019}: delocalized states are necessarily localized in reciprocal space. Our shifted grid technique can sample the reciprocal space finely enough to capture that feature. Olovsson \emph{et al.}\ on the other hand use  a $11\times11\times3$ $\bm{k}$-mesh. While this is sufficiently fine to reproduce the main features, it might not be sufficient to resolve the fine structure in sufficient detail.

\subsection{Hexagonal boron-nitride}

\begin{figure}
\centering
\includegraphics[scale=0.9]{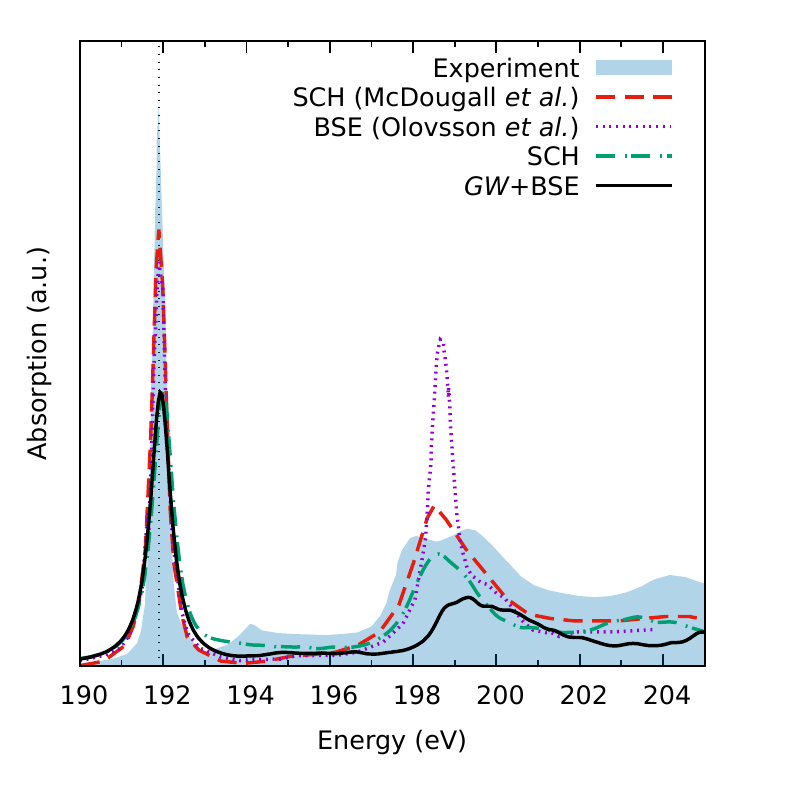}
\caption{XAS spectra of the $K$-edge of $B$ in $h$-BN. \emph{Blue filled curve:} experiment \cite{Li2012}. \emph{Black curve:} $GW$+BSE spectrum, this work.  \emph{Green dashed curve: } SCH-spectrum from \cite{McDougall2014}. \emph{Violet dotted curve:} BSE-spectrum from \cite{Olovsson2019}. \emph{Red dash-dotted curve:} SCH-spectrum. All modeled spectra are centered at $192$ eV and adjusted in height to match the integral over the first peak. }
\label{hBNB3}
\end{figure}
\begin{figure}
\centering
\includegraphics[scale=1]{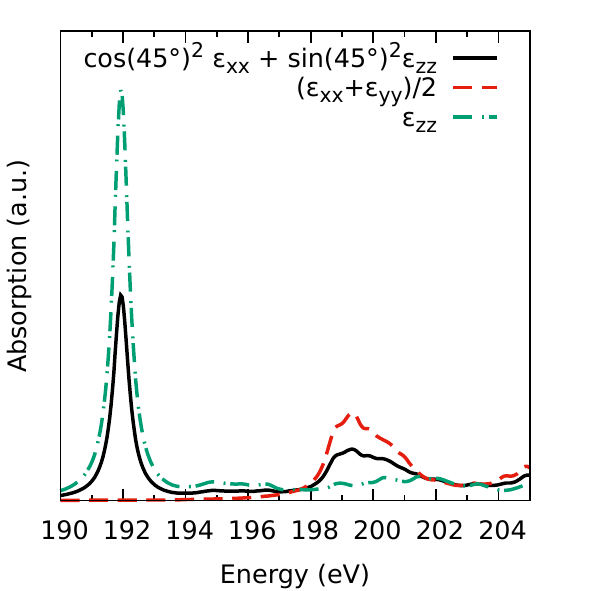}
\caption{Components of the dielectric tensor for the $K$-edge of $B$ in $h$-BN. \emph{Black curve:} components averaged according to $\text{Im}(\varepsilon) \sim \cos(45^\circ)^2 \varepsilon_{xx}+\sin(45^\circ)^2 \varepsilon_{zz}$. \emph{Orange dashed curve:} in-plane components, $\frac{1}{2}(\varepsilon_{xx}+\varepsilon_{yy})$. \emph{Green dash-dotted curve:} out-of-plane component, $\varepsilon_{zz}$.}
\label{hBNB2}
\end{figure}
\begin{figure}
\centering
\includegraphics[scale=1]{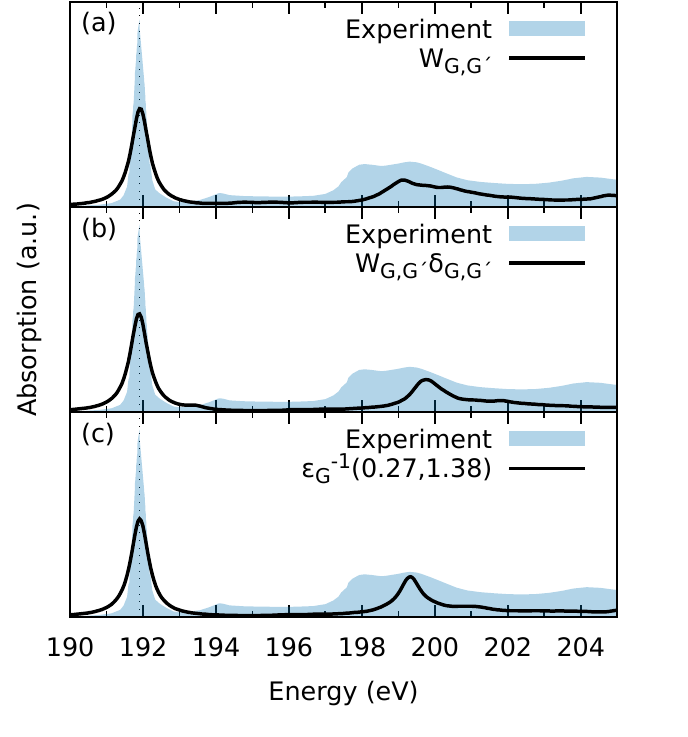}
\caption{$K$-edge of $B$ in $h$-BN, comparing different approaches to the screened interaction $W$. (a) $W$ from $G_0W_0$ step. (b) same as panel (a), excluding off-diagonal elements. (c) Model dielectric function  $\varepsilon_G^{-1}(0.27,1.38)$. \emph{Blue filled curve:} experiment \cite{Li2012}. } 
\label{hBNB1}
\end{figure}

\subsubsection{Boron}

In Fig.\ \ref{hBNB3} we compare an experimental $K$-edge spectrum of boron in $h$-BN  \cite{Li2012} to the XAS spectrum obtained with the present PAW--$GW$+BSE implementation and to an all-electron full-potential BSE spectrum \cite{Olovsson2019}. We also show two SCH spectra, one obtained with an all-electron APW implementation  \cite{McDougall2014} and an SCH spectrum calculated using the implementation presented previously \cite{Karsai2018}. All spectra have been adjusted in height such that the integrals over the first peak match. All methods reproduce the distance between the exciton peak at 192 eV and  the center of the double peak structure, located at 199 eV, while no spectrum can reproduce the double-peak structure or the smaller peak at 194 eV. As for graphite, the double peak structure has been attributed to a breaking of the degeneracy induced by the electron-phonon interaction \cite{Karsai2018,Olovsson2019}. Here our results are very clear: without the inclusion of electron-phonon couplings, one cannot reproduce the double-peak structure. We note that the experimental double peak is also much more pronounced and distinct than in graphite. 

As for graphite, we also show the in-plane and the out-of-plane components of the dielectric tensor in Fig.\ \ref{hBNB2}. Here we see that the strong peak at $192$ eV corresponds to the out-of-plane component, while  the features around $199$ eV are attributable to the in-plane components.

 In Fig.\ \ref{hBNB1} we compare the experiment to BSE spectra, where we use different approximations for the screened interaction, as outlined in the  section on computational methods and as explored for diamond before. We use either $W$ including (panel (a)) or excluding (panel (b)) off-diagonal elements or a model dielectric function $\varepsilon_G^{-1}(0.27,1.38)$ (panel (c)).  Here, only the $GW$+BSE spectrum including the off-diagonal elements (panel (a)) can reproduce the energy difference between the first peak at 192 eV and the middle of the double peak structure located at 199 eV. Comparing the model dielectric function approach to the approach in which only the diagonal elements of $W$ are kept we see that neither method can reproduce the position of the center of the double peak structure. However, we point out that the model-dielectric function approach performs slightly better, since in that case the second peak is redshifted by approximately 0.1 eV toward the center of the double peak at 199 eV.
 
\begin{figure}
\centering
\includegraphics[scale=0.9]{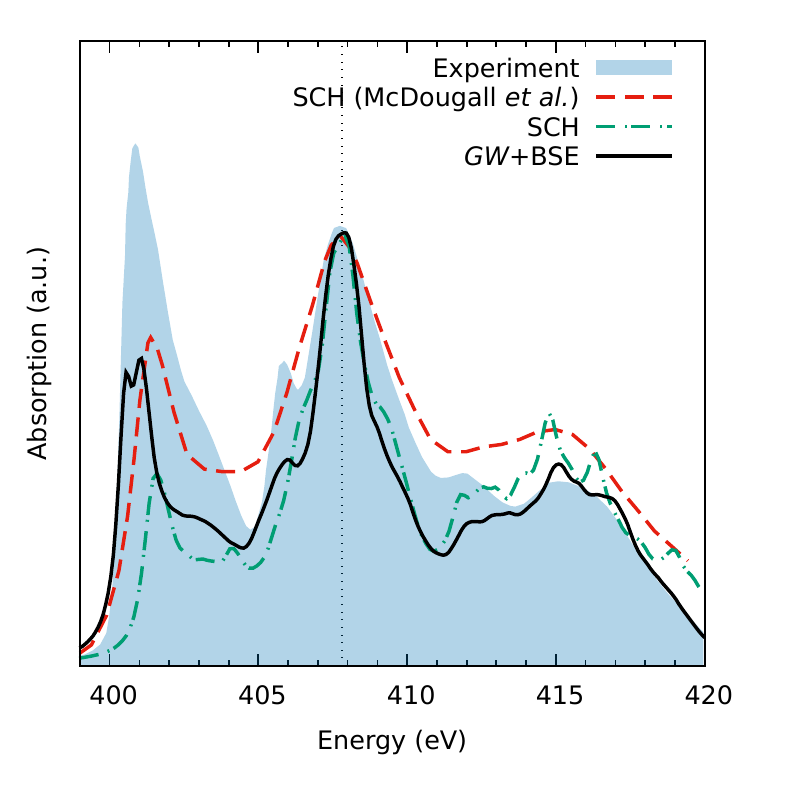}
\caption{XAS spectra of the $K$-edge of $N$ in $h$-BN. \emph{Blue filled curve:} experiment \cite{Petravic2013}. \emph{Black curve:} $GW$+BSE spectrum, this work. \emph{Green dashed curve:} SCH-spectrum  from \cite{McDougall2014}. \emph{Red dash-dotted curve:} SCH-spectrum.  All simulated spectra are centered on the maximum at  $408$ eV.  }
\label{hBNN1}
\end{figure}

\begin{figure}
\centering
\includegraphics[scale=1]{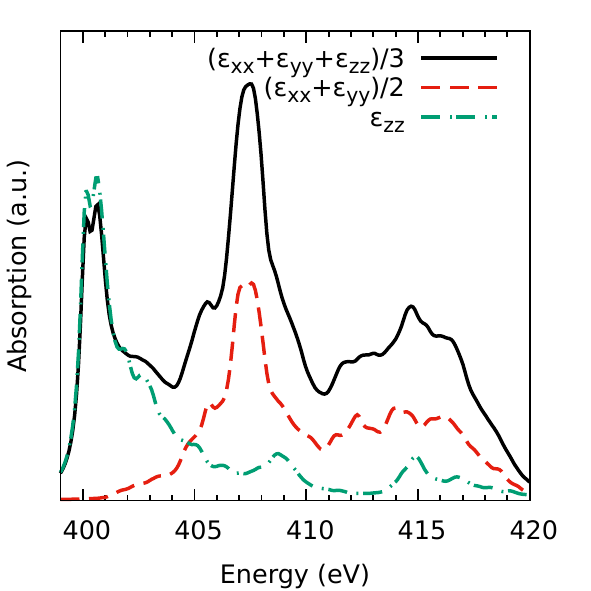}
\caption{Components of the dielectric tensor for the $K$-edge of $N$ in $h$-BN. \emph{Black curve:} average of all components, $\frac{1}{3}(\varepsilon_{xx}+\varepsilon_{yy}+\varepsilon_{zz})$. \emph{Orange dashed curve:} in-plane components, $\frac{1}{2}(\varepsilon_{xx}+\varepsilon_{yy})$. \emph{Green dash-dotted curve:} out-of-plane component, $\varepsilon_{zz}$.
}
\label{hBNN2}
\end{figure}

\subsubsection{Nitrogen} In Fig.\ \ref{hBNN1} we show an experimental $K$-edge spectrum of nitrogen in $h$-BN \cite{Li2012}, a modeled spectrum using the current PAW--$GW$+BSE implementation, an all-electron APW SCH-spectrum \cite{McDougall2014}, and an SCH-spectrum. Compared to the SCH spectrum by McDougall \emph{et al.}\ and the present SCH-spectrum, we are able to better reproduce the energy separation between the $\pi^*$ resonance at 401 eV and the $\sigma^*$ peak at 408 eV.  Furthermore, the $GW$+BSE spectrum shows a shoulder at 403 eV and a pronounced side-peak at 406 eV. These are also clearly visible in the experimental spectrum but absent in the SCH calculations. Finally, the shape of the first peak of the $GW$+BSE spectrum agrees better with experiment.

Again, we show the in-plane and out-of-plane components in Fig.\ \ref{hBNN2}. Analogously to graphite, we see that the first peak at 401 eV can be attributed to the out-of-plane component and is identified with the $\pi^*$ resonance.

\subsection{Lithium-halides}

\begin{figure}
\centering
\includegraphics[scale=0.9]{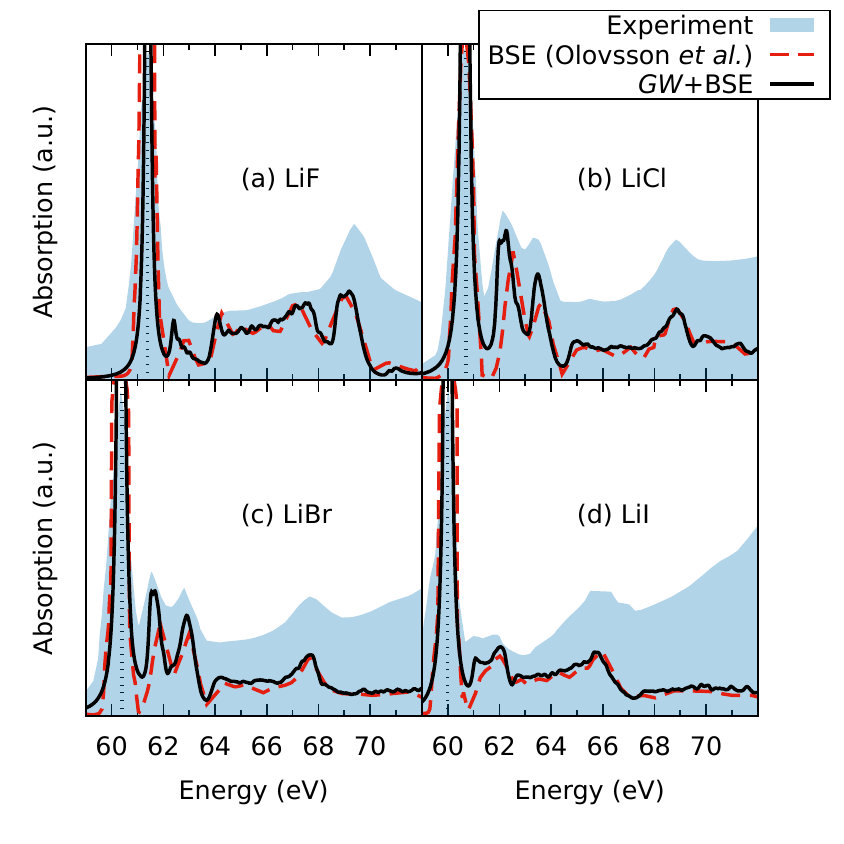}
\caption{XAS spectra for the $K$-edge of lithium in four lithium-halides. \emph{Blue filled curves:} experiment \cite{Handa2005}. \emph{Black curves:} $GW$+BSE spectra, this work. \emph{Red dashed lines:} BSE spectra from \cite{Olovsson2009}. All modeled spectra are centered on the vertical dashed line in each panel.}
\label{fig:halides}
\end{figure}

In Fig.\ \ref{fig:halides} we show the results for the XAS spectra of four lithium-halides with shallow core-states: LiF, LiCl, LiI, and LiBr. In each panel we compare our PAW--$GW$+BSE results to experiment \cite{Handa2005} and to an all-electron full-potential BSE spectrum \cite{Olovsson2009}. Both XAS-spectra of this work as well as of Olovsson \emph{et al.}\ can reproduce the edge-positions and the fine structure found in the experimental spectrum, however, our result can better match the peak positions of some features. This can be likely attributed to the fact that in the $GW$ step we calculate quasiparticle energies of each conduction band individually, while in the spectra of Olovsson \emph{et al.}\ the bands are shifted rigidly via a scissor operator. We note that we also used a much finer $\bm k$-point sampling, which can also change the shape and slightly the position of some peaks.

\section{Self-interaction effects in SCH calculations}\label{sec:error}

\begin{figure}
\centering
\includegraphics[width=\columnwidth]{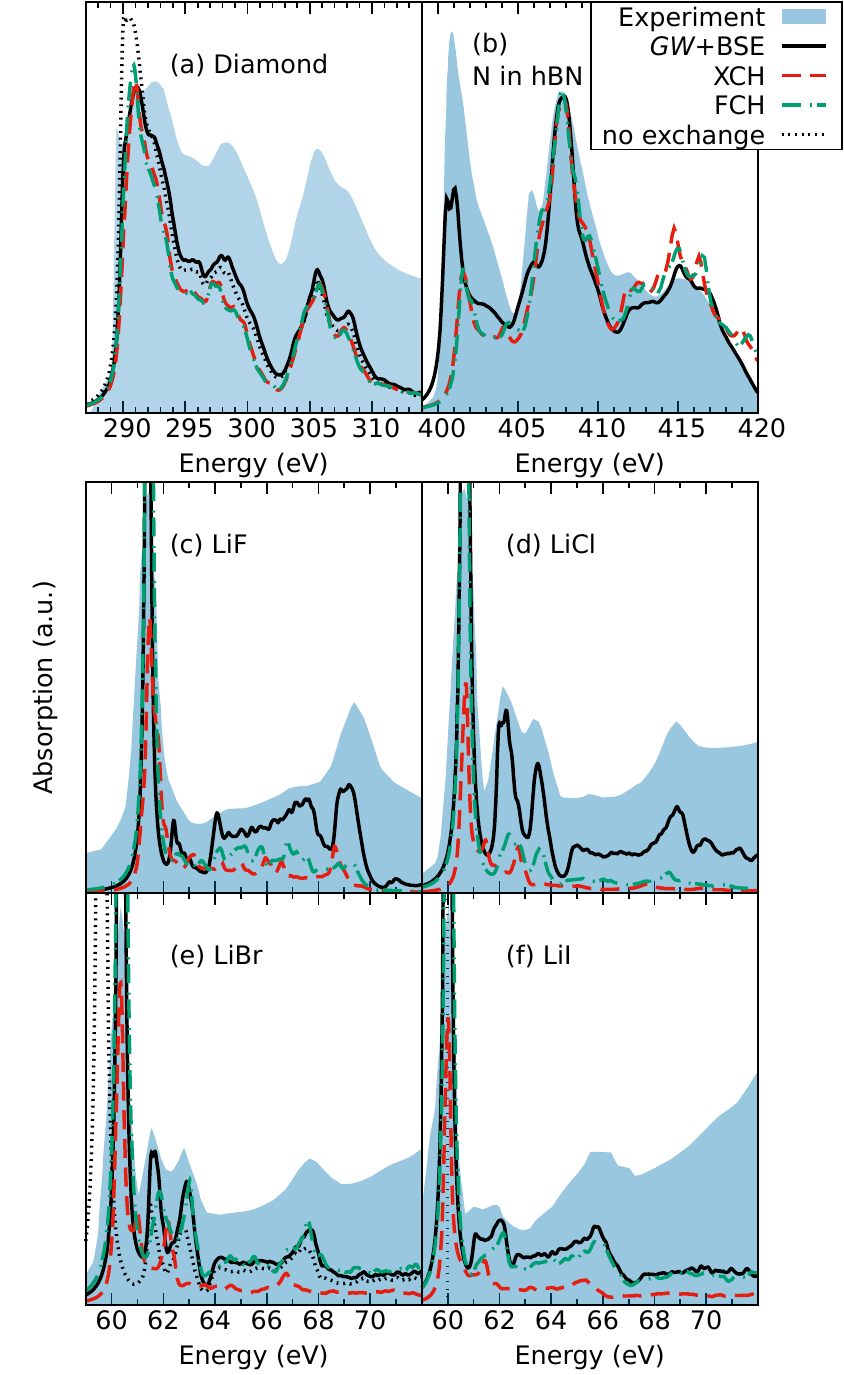}
\caption{XAS spectra for the $K$-edges of various systems. We show experimental spectra \cite{Handa2005,Ma1992, Petravic2013} (\emph{blue filled curves}), spectra using the current $GW$+BSE implementation (\emph{black curves}), XCH spectra where the electron was added to the lowest conduction band (\emph{red dashed curve}), and FCH spectra where the electron was added to the background charge (\emph{green dash-dotted curves}). In panels (a) and (e) we also show with a dotted line $GW$+BSE spectra without exchange term.}
\label{fig:improved_sch}
\end{figure}

The main issue that we will discuss in this paragraph is the dependence of the SCH method on where one places the excited electrons.
To this end, we collect calculated SCH spectra for diamond, N in $h$-BN, and lithium-halides in Fig.\ \ref{fig:improved_sch}. The technical details for the supercell calculations are collected in Table \ref{kpointstable}. 

We start with a brief discussion of the SCH method as it is commonly used in quantum chemistry and solid state physics. The most refined approach is to create a potentially fractional core-hole, and at the same time add a fractional charge to a conduction band state and perform a self-consistent DFT calculation. Ideally, self-consistent calculations should be performed for any of the many possible conduction band states. Then the transition probability from the groundstate into the excited state is calculated and the spectrum is obtained by combining all these calculations, where the excitation energy is given by the energy difference between the groundstate and the excited state and the amplitude by the transition probability. This requires many DFT calculations and is highly impractical for solid state calculations since it is impossible to place the electron in a selected conduction band if many $\bm{k}$-points are used. So in practice, in solid state calculations, the excited electrons are only placed into the conduction band edge or, even simpler, treated as negative background charge \cite{Hetenyi2004,Prendergast2006}.

Up to this point, all reported SCH spectra have been obtained by adding the core electron back to the conduction band edge, performing a single self-consistent DFT calculation and then calculating the transition probabilities into all conduction band states. In quantum chemistry, this approach is sometimes more specifically referred to as eXcited electron and Core-Hole method (XCH) \cite{Prendergast2006}. Placing the electron into the background charge, hence essentially omitting the excited electron density distribution in the self-consistent calculations, is sometimes referred to as the Full Core-Hole (FCH) method, since a full electron is removed from the core \cite{Hetenyi2004}.

At this point, it is expedient to study those two approaches, the XCH method, and the FCH method, for the materials scrutinized here. In the case of diamond and $h$-BN, panels (a) and (b) of Fig.\ \ref{fig:improved_sch} show that the  XCH and FCH spectra are almost identical --- compare red and green dashed lines. This implies that the electron that we place into the conduction band is well approximated by a homogeneous background charge, and this in turn  suggests that the localization of the excited electron is not very strong in the XCH method for diamond and $h$-BN.

However, this is not the case for the lithium-halides, Figs.\ \ref{fig:improved_sch}(c)-(f). Here we observe significant differences between placing the electron into the conduction band edge or into the background, again compare red and green line. Compared to the experimental spectrum and $GW$+BSE, the homogeneous background method (FCH, green line) yields clearly improved results for the peak positions compared to the more often used XCH method (red line). The peak positions for LiCl, LiBr, and LiI are in almost perfect agreement with the $GW$+BSE results, whereas for the XCH method the higher energy excitations are far too close to the main peak. Obviously, placing the excited electron into the lowest conduction band and then calculating the entire spectrum is not a good approximation for the lithium-halides, but works reasonably well for diamond and $h$-BN.

\begin{figure}[ht]
\includegraphics[width=\columnwidth]{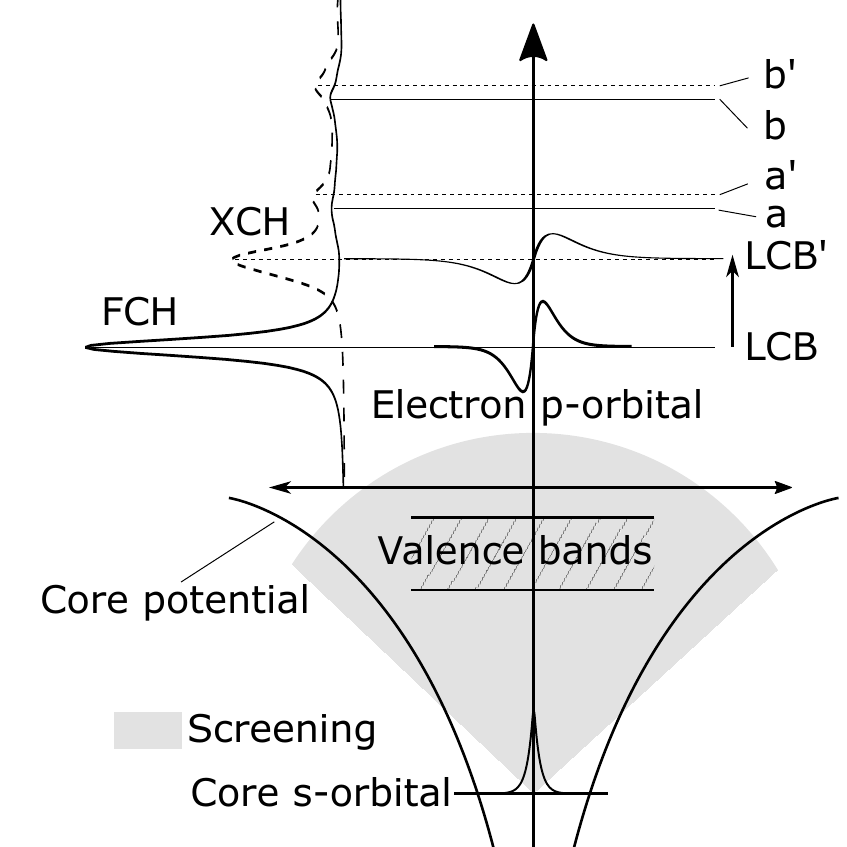}
\caption{Sketch describing how the self-interaction error of the additional electron in XCH method can change the spectra. See text for description.}
\label{fig:error}
\end{figure}

We will now try to explain why the SCH method often works well and when it tends to fail. 
The first crucial point is that the SCH method yields a reasonable approximation to the BSE, in particular, to the crucial term that describes the electrostatic interaction between the core-hole and the excited conduction band electrons. This is related to the $W$ term (direct term) in the BSE method. In the SCH method, a core-hole is created and all the other valence electrons will screen this local core-hole. The screening of the valence electrons is described by the static DFT dielectric function, so effectively the conduction band electrons see an effective screened core-hole $\int \varepsilon^{-1}(r',r) / |r-r_c|d^3 r $, if the core-hole is positioned at $r_c$. This implies that the SCH method mimics the effects of the direct term in the BSE very well. The successes and failures of the SCH method are then mainly related to the self-interaction of the conduction band electron. 

This will be discussed by inspecting the scheme shown in  Fig.\ \ref{fig:error}. The FCH spectrum of the $K$-edge of Li in Li-halides is schematically shown vertically in the left part of the figure. The lowest conduction band level is indicated by the line labeled \emph{LCB}. For the FCH method, this level is unoccupied, since the excited core electron is moved to the homogeneous background. To represent the orbital of the excited conduction band electron we show schematically a $p$-like orbital. In the XCH method, this orbital \emph{LCB}  becomes occupied with a single electron that essentially experiences the screened Coulomb potential (see above). The resulting localized charge density is added to the total charge density and in turn this further modifies the Hartree and DFT exchange-correlation potentials. Unless the DFT functional is cancelling the self-interaction error, this will lead to a sizeable self-interaction of the electron. In particular, this self-interaction error is more substantial, if the conduction band electron is strongly localized, as it is for the Li-halides. 

The self-interaction error in turn causes two effects: first, it shifts the energy level of the conduction band electron up from the initial energy to the new energy \emph{LCB'}. Second, it delocalizes the electron, which we have visualized by spreading out the $p$-orbital. These effects negatively impact the spectrum of the XCH method, as shown schematically in the left part of the figure. The delocalization reduces the oscillator strength, in turn reducing the amplitude of the first peak. The upshift of the eigenenergy on the other hand shifts the position of the peak to higher energies. Furthermore, this conduction band electron somewhat screens the core-hole. The energies of higher lying unoccupied conduction band states are then determined  in the presence of the screened core-hole potential and the potential of the added electron. This also shifts their energies upwards from $a,b$ to $a',b'$. However, since these states are necessarily orthogonal to the lowest conduction band state and since these states are also not as well localized as the lowest conduction band state, the energy shift $a\to a'$ is much smaller than from $\mathrm{LCB}\to\mathrm{LCB}'$. This negatively impacts the calculated spectrum and results in a too small energy separation between the main peak ($\mathrm{LCB}'$) and the other peaks ($a'$ and $b'$). 

To summarize, the main issue with the conventional XCH method is that the conduction band edge is shifted upwards due to self-interaction errors present in most semi-local DFT functionals. If the additional electron is placed into the background this problem is not observed. Note that self-interaction is unphysical and would not be present for the exact DFT functional, and one might well argue that the conventional XCH methods in combination with DFT functionals that are not self-interaction free is bound to be fairly inaccurate for localized excitons.

To make a connection to the $GW$+BSE approach and to further substantiate our claims we have also investigated how the exchange terms contribute to the $GW$+BSE spectra. The exchange term has a positive sign and is related to the change of the Hartree potential. It accounts for the repulsive electrostatic interaction of the individual electron-hole pairs, see Eq.\ \eqref{eq:exc_matrixelements}. If the electron-hole pairs are strongly localized, then including the exchange term will shift oscillator strengths to higher energies. Conversely, omitting the exchange term for strongly localized electron-hole pairs shifts the spectrum to lower energies. Now, above we argued that self-interaction Hartree effects come into play for strongly localized excitons. Hence, for excitons involving conduction band states where the Hartree self-interaction is large, we should also see a noticeable energy-shift of the excitonic peaks when the exchange term is omitted.

In panel (a) and (e) of Fig.\ \ref{fig:improved_sch}, we show the $GW$+BSE results for diamond and Li in LiBr where the  exchange terms was omitted (black dashed line). For diamond this hardly changes the spectrum, in agreement with the observation that the XCH and FCH method are pretty much identical. For LiBr though, the first peak, and the first peak only, is shifted towards the left by almost 1 eV. This confirms that the excitation into the first conduction band state in LiBr creates a strongly localized exciton, whereas all the other excitations in LiBr as well as in diamond are fairly delocalized, in line with our previous arguments. 

A final note is in place: in the  BSE method changes of the Hartree potential are considered for all excitations individually, whereas in the XCH method one places the electron into the lowest conduction band and assumes that changes of the Hartree potential for this case mimic the changes of the Hartree potential, if one would place the electrons into other conduction band states. This is obviously a bad approximation, if the lowest conduction band state is strongly localized whereas the other states are delocalized. It seems wiser to  leave changes of the potential related to the occupation of the lowest conduction band entirely out, as done for the FCH method.

\section{Summary and Conclusion}\label{sec:summary}

In the present work we have discussed how to implement the BSE within the PAW methodology for the calculation of X-ray absorption spectra. After reviewing the basic theory, we have detailed how to evaluate the matrix elements of the BSE Hamiltonian and explained how the expressions need to be modified when core states are included. In particular, we have discussed that we approximate the overlap charge densities involving core states using augmentation charges only. This suffices, since
the augmentation charges are designed to describe long-range electrostatic interactions essentially exactly. However, for the transition probabilities, we use the exact all-electron core orbitals and the all-electron partial waves corresponding
to the conduction band electrons. We have tested our implementation on four materials classes: a covalent system (diamond), two 2D van-der-Waals layered materials (graphite and $h$-BN) and four systems with shallow core states (Li-halides).

For diamond we found that our result could reproduce all relevant peak positions very well. Compared to other theoretical spectra our spectrum features an additional shoulder at around 290 eV, which we assign to the excitonic peak of the experimental spectrum. As was the case for the supercell core-hole method and previous BSE calculations, our spectrum lacks intensity at higher energies. This seems to be mostly related to a background that is present in the experimental data but missing at the level of theory that is commonly used. The origin of the lack of background in the present theories, is not entirely clear to us, but could be related to interactions with core-conduction band pairs at other sites. 

For diamond we have also investigated the influence of different approximations to the screened interaction $W$ on the spectrum. We found that the  off-diagonal elements of the screened interaction play an important role for accurate predictions of the individual peaks in the core-hole spectrum; neglecting the off diagonal components leads to a quite significant deterioration of the results compared to experiment. This also means that simplified approaches that attempt to model the screening using a diagonal model lead to inaccurate results. 

In graphite the peak positions and intensity ratios also match the experiments well. In our results, we found a shoulder in the $\sigma^*$ peak that can be tentatively assigned to the $\sigma_2^*$ peak, suggested by some authors to be related to a delocalized bandlike contribution \cite{Bruehwiler1995}. So strictly speaking, electron-phonon coupling may not be required to explain the splitting of the $\sigma^*$ peak, but we also do not doubt that the splitting will be enhanced by electron-phonon coupling, as suggested by other authors \cite{Olovsson2019}.

For the boron $K$-edge in $h$-BN we found again good agreement with experiment and previous BSE and supercell calculations. The results for the nitrogen $K$-edge are somewhat more notable. Previous supercell core-hole calculations lack important features in the spectrum, and the $GW$+BSE approach is able to describe all relevant peaks observed experimentally. Specifically, the distance between the  $\pi^*$ and $\sigma^*$ peaks is in excellent agreement with experiment, and the $GW$+BSE can also  reproduce experimental peaks previously missing in the simulations.

For the Li-halides we found excellent agreement between the BSE results and experiment. Compared to previous  BSE spectra employing a scissor shift, a slight but noticeable improvement of the peak positions is observed. We have tentatively related this to improved quasiparticle energies, as our calculations rely on the $G_0W_0$  band structures, whereas previous calculations only employed a scissor corrected DFT band structure. We also note that our $\bm{k}$-point sampling was vastly improved over previous work, which potentially has also positively impacted our predicted spectra.

Finally, we have carefully scrutinized the supercell core-hole method for diamond, $h$-BN, as well as the Li-halides. As discussed, the supercell core-hole method comes in two variants, one where the excited electron is placed in the conduction band edge, and the other  where the electron is treated as a homogeneous background charge.
For diamond and $h$-BN, both approaches yield pretty much identical results. For the  halides, treating the excited electron as a background charge gives much improved peak positions almost on par with the BSE. We have tried to argue that the self-interaction errors of present density functionals are the most likely explanation for the failure of the supercell core-hole method with the excited electron in the lowest conduction band. Self-interaction erroneously reduces the oscillator strength and pushes the first excitonic peak to too high energies.

In summary, the present work shows that excellent predictions for core-hole spectra are possible using the PAW method and the BSE approach. Since the PAW method can be routinely applied to fairly large systems, and since the existing BSE implementation also works for several hundred thousand electron-hole pairs, applications to reasonably large structures are now possible on a fairly routine basis.

\section*{Acknowledgements}

This research was funded by the Austrian Science Fund (FWF) DOC 85-N.

\section*{Appendix: construction of augmentation charges}
The construction of the augmentation charges proceeds in two steps: first moment restoration and then shape restoration.

We start with moment restoration and write down the defining equation for the augmentation charges, the requirement that the augmentation charges have the same moments as the difference of the AE and PS charge densities inside the PAW sphere :
\begin{equation}\label{augemntationcharge_definition}
    \int_{\Omega_r} [n^1(\bm{r}) - \tilde n^1(\bm{r}) - \hat n(\bm{r})] |\bm{r}|^L Y^*_{LM}(\Omega)  d^3 r=0,
\end{equation}
where $Y^*_{LM}(\Omega)$ are the spherical harmonics, $L$ and $M$ the total orbital and magnetic quantum numbers, and $\Omega=\{\theta,\phi\}$ the angular variables. In the PAW method the charge density difference $n^1(\bm{r}) - \tilde n^1(\bm{r})$ is written as \cite{Kresse1999}
\begin{equation}\label{difference}
 n^1(\bm{r}) - \tilde n^1(\bm{r}) = \sum_{nm} \rho_{nm} Q_{nm}(\bm{r}),
\end{equation}
with the functions 
\begin{equation}\label{charge_diff}
 Q_{nm}(\bm{r}) = \phi_n^*(\bm{r}) \phi_m(\bm{r}) -\tilde\phi_n^*(\bm{r}) \tilde\phi_m(\bm{r})
\end{equation}
and occupancies (or one-center density matrix)
\begin{equation}
    \rho_{nm} = \langle\tilde\psi| \tilde p_n \rangle \langle \tilde p_m |\tilde\psi\rangle.
\end{equation}
Furthermore, the indices $n,m$ are compound indices: $n=\{\epsilon_n,l_n,m_n\}$.

We now calculate the moments of the charge difference:
\begin{equation}
    q_{nm}^{LM} =  \int_{\Omega_r} Q_{nm}(\bm{r}) |\bm{r}|^L Y^*_{LM}(\Omega)  r^2dr d\Omega.
\end{equation}
In the PAW formalism, AE and PS partial waves, Eqs.\ \eqref{AE_partial} and \eqref{PS_partial}, respectively, are products of radial waves and spherical harmonics. Inserting these into the preceding equation, the integral separates into radial and angular parts
\begin{equation}\label{moments}
\begin{split}
        q_{nm}^{LM} &=  \int_0^{r_c} (u_{k_n,l_n}u_{k_m,l_m}-\tilde u_{k_n,l_n}\tilde u_{k_m,l_m})|\bm{r}|^L  dr \\
        &\times\int Y_{l_n,m_n}(\Omega) Y_{l_m,m_m}(\Omega) Y^*_{LM}(\Omega)d\Omega.
\end{split}
\end{equation}
The radial integral of three spherical harmonic (also called Gaunt coefficients) imposes  conditions on the angular momenta $l_n,l_m$, and $L$. In particular, all the usual rules of the addition of angular momentum apply: $m_n+m_m=M$ and
\begin{equation}
    L=|l_n-l_m|,|l_n-l_m|+2,\dots,|l_n+l_m|.
\end{equation}

After having calculated the moments, Eq.\ \eqref{moments}, the augmentation charge is
\begin{equation}
    \hat n = \sum_{\substack{LM\\l_n,m_n,l_m,m_m\\n,m} } \rho_{nm} q_{nm}^{LM}Y_{LM}(\Omega).
\end{equation}
The one-center density matrix $\rho_{nm}$ is indexed by the angular momenta $l_n,m_n,l_m,m_m$. On the other hand, the augmentation charge that will be added to the plane wave grid is indexed by the total orbital and magnetic quantum numbers $L,M$. We therefore need a change of basis to a $\rho_{LM}$ density matrix. To achieve this we sum over the $l,m$ indices in the preceding equation. This results in the required $L,M$-dependent density matrix $\rho_{LM}$ and the $L,M$-dependent moments $q^{LM}_{nm}$:
\begin{equation}\label{augmentation_charge_result}
    \sum_{l_n,m_n,l_m,m_m} q_{nm}^{LM} \rho_{l_n,m_n,l_m,m_m}= q^{LM}_{nm} \rho_{LM}.
\end{equation}
We obtain then our final result for the moment-restoring augmentation charges:
\begin{align}
    \hat n &= \sum_{\substack{LM\\n,m}} \rho_{LM} \hat Q^{LM}_{nm},\\
    \hat Q^{LM}_{nm} &= q^{LM}_{nm}  Y_{LM}.
\end{align}
These augmentation charges are added to the PS charge density on the plane-wave grid, as in equation \eqref{charge_approx}. We note that the augmentation charges are directly added in real space.

 Up to this point we neglected one-center terms and reconstructed the augmentation charges from the moments of the exact charge density inside the PAW sphere.  For post-DFT methods it can be beneficial to approximate the contributions of the one-center terms and to add these contributions to the augmentation charges. This process is called shape restoration. We add to the right hand side of equation \eqref{augmentation_charge_result} for each total angular momentum $L$ shape-restoring radial functions $\Delta g_{L}(r)$, with coefficients $c^{L}_{nm} $ to be determined:
\begin{equation}
 \hat n = \sum_{\substack{LM\\n,m}} \rho_{LM} \left[ q^{LM}_{nm} Y_{LM}(\Omega) + c^{L}_{nm}  \Delta g_{L}(r) Y_{LM}(\Omega)\right],
\end{equation}
where the shape-restoring functions are written as a sum of two spherical Bessel-functions:
\begin{equation}
    \Delta g_{L}(r) = \sum_{\beta=1}^2\alpha_\beta^L j_L(q_\beta^L r).
\end{equation}
The coefficients $\alpha_\beta^L$ and $q_\beta^L$ are chosen such that the $l,m$-multipole of the shape restoring charge contribution vanishes and that the  Hankel transforms of $\Delta g_{L}(r)$ and $Q_{nm}(r)$ 
\begin{align}
    Q_{nm}(q)&=\int_0^{R_c} Q_{nm}(r) j_L(q r) r^{L} dr,\\
    \Delta g_{L}(q)&=\int_0^{R_c} \Delta g_{L}(r)j_L(q r) r^{L} dr\\
\end{align}
are identical at chosen values of $q$.

To calculate the coefficients $c^{L}_{nm}$ we subtract from the radial part of the charge density difference, Eq.\ \eqref{charge_diff}, the radial part of the moment-restoring contribution to the augmentation charge:
\begin{equation}
    \delta Q_{nm}(r)= Q_{nm}(r)-\hat Q_{nm}(r).
\end{equation}
In the next step we write the quantity $\delta Q(r)$ as a superposition of shape restoring functions:
\begin{equation}
    \delta Q_{nm}(r)=\sum_{L}c^{L}_{nm}\Delta g_{L}(r). 
\end{equation}
We then multiply the preceding equation by $j_L(q_\beta^L r)$ and integrate over $r^L$:
\begin{equation}\label{pre_matrix_form}
\begin{split}
    \int j_L(q_\beta^L r)\delta Q_{nm}(r) r^L dr &=\\
    \sum_{L}c^{L}_{nm}&\int j_l(q_\beta^L r) \Delta g_{L}(r) r^L dr.
\end{split}
\end{equation}
Introducing coefficients
\begin{align}
    b_\beta&=\int j_L(q_\beta^L r)\delta Q_{nm}(r)r^L dr,\\
    A_{\beta,L} &=\int j_L(q_\beta^L r) \Delta g_{L}(r)r^L dr,
\end{align}
equation \eqref{pre_matrix_form} can be recast as a system of linear equations:
\begin{equation}
    \sum_L A_{\beta,L} c^{L}_{nm}=b_\beta,
\end{equation}
which can be solved by standard methods to determine the coefficients $c^{L}_{nm}$.

\bibliography{XAS_bibliography} 

\begin{thebibliography}{49}%
\makeatletter
\providecommand \@ifxundefined [1]{%
 \@ifx{#1\undefined}
}%
\providecommand \@ifnum [1]{%
 \ifnum #1\expandafter \@firstoftwo
 \else \expandafter \@secondoftwo
 \fi
}%
\providecommand \@ifx [1]{%
 \ifx #1\expandafter \@firstoftwo
 \else \expandafter \@secondoftwo
 \fi
}%
\providecommand \natexlab [1]{#1}%
\providecommand \enquote  [1]{``#1''}%
\providecommand \bibnamefont  [1]{#1}%
\providecommand \bibfnamefont [1]{#1}%
\providecommand \citenamefont [1]{#1}%
\providecommand \href@noop [0]{\@secondoftwo}%
\providecommand \href [0]{\begingroup \@sanitize@url \@href}%
\providecommand \@href[1]{\@@startlink{#1}\@@href}%
\providecommand \@@href[1]{\endgroup#1\@@endlink}%
\providecommand \@sanitize@url [0]{\catcode `\\12\catcode `\$12\catcode
  `\&12\catcode `\#12\catcode `\^12\catcode `\_12\catcode `\%12\relax}%
\providecommand \@@startlink[1]{}%
\providecommand \@@endlink[0]{}%
\providecommand \url  [0]{\begingroup\@sanitize@url \@url }%
\providecommand \@url [1]{\endgroup\@href {#1}{\urlprefix }}%
\providecommand \urlprefix  [0]{URL }%
\providecommand \Eprint [0]{\href }%
\providecommand \doibase [0]{https://doi.org/}%
\providecommand \selectlanguage [0]{\@gobble}%
\providecommand \bibinfo  [0]{\@secondoftwo}%
\providecommand \bibfield  [0]{\@secondoftwo}%
\providecommand \translation [1]{[#1]}%
\providecommand \BibitemOpen [0]{}%
\providecommand \bibitemStop [0]{}%
\providecommand \bibitemNoStop [0]{.\EOS\space}%
\providecommand \EOS [0]{\spacefactor3000\relax}%
\providecommand \BibitemShut  [1]{\csname bibitem#1\endcsname}%
\let\auto@bib@innerbib\@empty
\bibitem [{\citenamefont {Yano}\ and\ \citenamefont
  {Yachandra}(2009)}]{Yano2009}%
  \BibitemOpen
  \bibfield  {author} {\bibinfo {author} {\bibfnamefont {J.}~\bibnamefont
  {Yano}}\ and\ \bibinfo {author} {\bibfnamefont {V.~K.}\ \bibnamefont
  {Yachandra}},\ }\href {https://doi.org/10.1007/s11120-009-9473-8} {\bibfield
  {journal} {\bibinfo  {journal} {Photosynth. Res.}\ }\textbf {\bibinfo
  {volume} {102}},\ \bibinfo {pages} {241} (\bibinfo {year}
  {2009})}\BibitemShut {NoStop}%
\bibitem [{\citenamefont {Hohenberg}\ and\ \citenamefont
  {Kohn}(1964)}]{Hohenberg1964}%
  \BibitemOpen
  \bibfield  {author} {\bibinfo {author} {\bibfnamefont {P.}~\bibnamefont
  {Hohenberg}}\ and\ \bibinfo {author} {\bibfnamefont {W.}~\bibnamefont
  {Kohn}},\ }\href {https://doi.org/10.1103/physrev.136.b864} {\bibfield
  {journal} {\bibinfo  {journal} {Phys. Rev.}\ }\textbf {\bibinfo {volume}
  {136}},\ \bibinfo {pages} {B864} (\bibinfo {year} {1964})}\BibitemShut
  {NoStop}%
\bibitem [{\citenamefont {Onida}\ \emph {et~al.}(2002)\citenamefont {Onida},
  \citenamefont {Reining},\ and\ \citenamefont {Rubio}}]{Onida2002}%
  \BibitemOpen
  \bibfield  {author} {\bibinfo {author} {\bibfnamefont {G.}~\bibnamefont
  {Onida}}, \bibinfo {author} {\bibfnamefont {L.}~\bibnamefont {Reining}},\
  and\ \bibinfo {author} {\bibfnamefont {A.}~\bibnamefont {Rubio}},\ }\href
  {https://doi.org/10.1103/revmodphys.74.601} {\bibfield  {journal} {\bibinfo
  {journal} {Rev. Mod. Phys.}\ }\textbf {\bibinfo {volume} {74}},\ \bibinfo
  {pages} {601} (\bibinfo {year} {2002})}\BibitemShut {NoStop}%
\bibitem [{\citenamefont {Hybertsen}\ and\ \citenamefont
  {Louie}(1986)}]{Hybertsen1986}%
  \BibitemOpen
  \bibfield  {author} {\bibinfo {author} {\bibfnamefont {M.~S.}\ \bibnamefont
  {Hybertsen}}\ and\ \bibinfo {author} {\bibfnamefont {S.~G.}\ \bibnamefont
  {Louie}},\ }\href {https://doi.org/10.1103/physrevb.34.5390} {\bibfield
  {journal} {\bibinfo  {journal} {Phys. Rev. B}\ }\textbf {\bibinfo {volume}
  {34}},\ \bibinfo {pages} {5390} (\bibinfo {year} {1986})}\BibitemShut
  {NoStop}%
\bibitem [{\citenamefont {Sham}\ and\ \citenamefont {Rice}(1966)}]{Sham1966}%
  \BibitemOpen
  \bibfield  {author} {\bibinfo {author} {\bibfnamefont {L.~J.}\ \bibnamefont
  {Sham}}\ and\ \bibinfo {author} {\bibfnamefont {T.~M.}\ \bibnamefont
  {Rice}},\ }\href {https://doi.org/10.1103/PhysRev.144.708} {\bibfield
  {journal} {\bibinfo  {journal} {Phys. Rev.}\ }\textbf {\bibinfo {volume}
  {144}},\ \bibinfo {pages} {708} (\bibinfo {year} {1966})}\BibitemShut
  {NoStop}%
\bibitem [{\citenamefont {Martin}\ \emph {et~al.}(2016)\citenamefont {Martin},
  \citenamefont {Reining},\ and\ \citenamefont {Ceperley}}]{Martin2016}%
  \BibitemOpen
  \bibfield  {author} {\bibinfo {author} {\bibfnamefont {R.~M.}\ \bibnamefont
  {Martin}}, \bibinfo {author} {\bibfnamefont {L.}~\bibnamefont {Reining}},\
  and\ \bibinfo {author} {\bibfnamefont {D.~M.}\ \bibnamefont {Ceperley}},\
  }\href
  {https://www.ebook.de/de/product/25518288/richard_m_martin_lucia_reining_david_m_ceperley_interacting_electrons.html}
  {\emph {\bibinfo {title} {Interacting Electrons}}}\ (\bibinfo  {publisher}
  {Cambridge University Press},\ \bibinfo {year} {2016})\BibitemShut {NoStop}%
\bibitem [{\citenamefont {Hanke}\ and\ \citenamefont {Sham}(1979)}]{Hanke1979}%
  \BibitemOpen
  \bibfield  {author} {\bibinfo {author} {\bibfnamefont {W.}~\bibnamefont
  {Hanke}}\ and\ \bibinfo {author} {\bibfnamefont {L.~J.}\ \bibnamefont
  {Sham}},\ }\href {https://doi.org/10.1103/physrevlett.43.387} {\bibfield
  {journal} {\bibinfo  {journal} {Phys. Rev. Lett.}\ }\textbf {\bibinfo
  {volume} {43}},\ \bibinfo {pages} {387} (\bibinfo {year} {1979})}\BibitemShut
  {NoStop}%
\bibitem [{\citenamefont {Benedict}\ \emph {et~al.}(1998)\citenamefont
  {Benedict}, \citenamefont {Shirley},\ and\ \citenamefont
  {Bohn}}]{Benedict1998}%
  \BibitemOpen
  \bibfield  {author} {\bibinfo {author} {\bibfnamefont {L.~X.}\ \bibnamefont
  {Benedict}}, \bibinfo {author} {\bibfnamefont {E.~L.}\ \bibnamefont
  {Shirley}},\ and\ \bibinfo {author} {\bibfnamefont {R.~B.}\ \bibnamefont
  {Bohn}},\ }\href {https://doi.org/10.1103/PhysRevLett.80.4514} {\bibfield
  {journal} {\bibinfo  {journal} {Phys. Rev. Lett.}\ }\textbf {\bibinfo
  {volume} {80}},\ \bibinfo {pages} {4514} (\bibinfo {year}
  {1998})}\BibitemShut {NoStop}%
\bibitem [{\citenamefont {Albrecht}\ \emph {et~al.}(1998)\citenamefont
  {Albrecht}, \citenamefont {Reining}, \citenamefont {Del~Sole},\ and\
  \citenamefont {Onida}}]{Albrecht1998}%
  \BibitemOpen
  \bibfield  {author} {\bibinfo {author} {\bibfnamefont {S.}~\bibnamefont
  {Albrecht}}, \bibinfo {author} {\bibfnamefont {L.}~\bibnamefont {Reining}},
  \bibinfo {author} {\bibfnamefont {R.}~\bibnamefont {Del~Sole}},\ and\
  \bibinfo {author} {\bibfnamefont {G.}~\bibnamefont {Onida}},\ }\href
  {https://doi.org/10.1103/PhysRevLett.80.4510} {\bibfield  {journal} {\bibinfo
   {journal} {Phys. Rev. Lett.}\ }\textbf {\bibinfo {volume} {80}},\ \bibinfo
  {pages} {4510} (\bibinfo {year} {1998})}\BibitemShut {NoStop}%
\bibitem [{\citenamefont {van~der Horst}\ \emph {et~al.}(1999)\citenamefont
  {van~der Horst}, \citenamefont {Bobbert}, \citenamefont {Michels},
  \citenamefont {Brocks},\ and\ \citenamefont {Kelly}}]{Horst1999}%
  \BibitemOpen
  \bibfield  {author} {\bibinfo {author} {\bibfnamefont {J.-W.}\ \bibnamefont
  {van~der Horst}}, \bibinfo {author} {\bibfnamefont {P.~A.}\ \bibnamefont
  {Bobbert}}, \bibinfo {author} {\bibfnamefont {M.~A.~J.}\ \bibnamefont
  {Michels}}, \bibinfo {author} {\bibfnamefont {G.}~\bibnamefont {Brocks}},\
  and\ \bibinfo {author} {\bibfnamefont {P.~J.}\ \bibnamefont {Kelly}},\ }\href
  {https://doi.org/10.1103/PhysRevLett.83.4413} {\bibfield  {journal} {\bibinfo
   {journal} {Phys. Rev. Lett.}\ }\textbf {\bibinfo {volume} {83}},\ \bibinfo
  {pages} {4413} (\bibinfo {year} {1999})}\BibitemShut {NoStop}%
\bibitem [{\citenamefont {Rohlfing}\ and\ \citenamefont
  {Louie}(2000)}]{Rohlfing2000}%
  \BibitemOpen
  \bibfield  {author} {\bibinfo {author} {\bibfnamefont {M.}~\bibnamefont
  {Rohlfing}}\ and\ \bibinfo {author} {\bibfnamefont {S.~G.}\ \bibnamefont
  {Louie}},\ }\href {https://doi.org/10.1103/PhysRevB.62.4927} {\bibfield
  {journal} {\bibinfo  {journal} {Phys. Rev. B}\ }\textbf {\bibinfo {volume}
  {62}},\ \bibinfo {pages} {4927} (\bibinfo {year} {2000})}\BibitemShut
  {NoStop}%
\bibitem [{\citenamefont {Shirley}(1998)}]{Shirley1998}%
  \BibitemOpen
  \bibfield  {author} {\bibinfo {author} {\bibfnamefont {E.~L.}\ \bibnamefont
  {Shirley}},\ }\href {https://doi.org/10.1103/PhysRevLett.80.794} {\bibfield
  {journal} {\bibinfo  {journal} {Phys. Rev. Lett.}\ }\textbf {\bibinfo
  {volume} {80}},\ \bibinfo {pages} {794} (\bibinfo {year} {1998})}\BibitemShut
  {NoStop}%
\bibitem [{\citenamefont {Soininen}\ and\ \citenamefont
  {Shirley}(2001)}]{Soininen2001}%
  \BibitemOpen
  \bibfield  {author} {\bibinfo {author} {\bibfnamefont {J.~A.}\ \bibnamefont
  {Soininen}}\ and\ \bibinfo {author} {\bibfnamefont {E.~L.}\ \bibnamefont
  {Shirley}},\ }\href {https://doi.org/10.1103/PhysRevB.64.165112} {\bibfield
  {journal} {\bibinfo  {journal} {Phys. Rev. B}\ }\textbf {\bibinfo {volume}
  {64}},\ \bibinfo {pages} {165112} (\bibinfo {year} {2001})}\BibitemShut
  {NoStop}%
\bibitem [{\citenamefont {Vinson}\ \emph {et~al.}(2011)\citenamefont {Vinson},
  \citenamefont {Rehr}, \citenamefont {Kas},\ and\ \citenamefont
  {Shirley}}]{Vinson2011}%
  \BibitemOpen
  \bibfield  {author} {\bibinfo {author} {\bibfnamefont {J.}~\bibnamefont
  {Vinson}}, \bibinfo {author} {\bibfnamefont {J.~J.}\ \bibnamefont {Rehr}},
  \bibinfo {author} {\bibfnamefont {J.~J.}\ \bibnamefont {Kas}},\ and\ \bibinfo
  {author} {\bibfnamefont {E.~L.}\ \bibnamefont {Shirley}},\ }\href
  {https://doi.org/10.1103/PhysRevB.83.115106} {\bibfield  {journal} {\bibinfo
  {journal} {Phys. Rev. B}\ }\textbf {\bibinfo {volume} {83}},\ \bibinfo
  {pages} {115106} (\bibinfo {year} {2011})}\BibitemShut {NoStop}%
\bibitem [{\citenamefont {Gilmore}\ \emph {et~al.}(2015)\citenamefont
  {Gilmore}, \citenamefont {Vinson}, \citenamefont {Shirley}, \citenamefont
  {Prendergast}, \citenamefont {Pemmaraju}, \citenamefont {Kas}, \citenamefont
  {Vila},\ and\ \citenamefont {Rehr}}]{Gilmore2015}%
  \BibitemOpen
  \bibfield  {author} {\bibinfo {author} {\bibfnamefont {K.}~\bibnamefont
  {Gilmore}}, \bibinfo {author} {\bibfnamefont {J.}~\bibnamefont {Vinson}},
  \bibinfo {author} {\bibfnamefont {E.~L.}\ \bibnamefont {Shirley}}, \bibinfo
  {author} {\bibfnamefont {D.}~\bibnamefont {Prendergast}}, \bibinfo {author}
  {\bibfnamefont {C.~D.}\ \bibnamefont {Pemmaraju}}, \bibinfo {author}
  {\bibfnamefont {J.~J.}\ \bibnamefont {Kas}}, \bibinfo {author} {\bibfnamefont
  {F.~D.}\ \bibnamefont {Vila}},\ and\ \bibinfo {author} {\bibfnamefont
  {J.~J.}\ \bibnamefont {Rehr}},\ }\href
  {https://doi.org/10.1016/j.cpc.2015.08.014} {\bibfield  {journal} {\bibinfo
  {journal} {Comput. Phys. Commun.}\ }\textbf {\bibinfo {volume} {197}},\
  \bibinfo {pages} {109} (\bibinfo {year} {2015})}\BibitemShut {NoStop}%
\bibitem [{\citenamefont {Shirley}\ \emph {et~al.}(2020)\citenamefont
  {Shirley}, \citenamefont {Vinson},\ and\ \citenamefont
  {Gilmore}}]{Shirley2020}%
  \BibitemOpen
  \bibfield  {author} {\bibinfo {author} {\bibfnamefont {E.~L.}\ \bibnamefont
  {Shirley}}, \bibinfo {author} {\bibfnamefont {J.}~\bibnamefont {Vinson}},\
  and\ \bibinfo {author} {\bibfnamefont {K.}~\bibnamefont {Gilmore}},\ }in\
  \href {https://doi.org/10.1107/s157487072000333x} {\emph {\bibinfo
  {booktitle} {International Tables for Crystallography}}}\ (\bibinfo
  {publisher} {International Union of Crystallography},\ \bibinfo {year}
  {2020})\BibitemShut {NoStop}%
\bibitem [{\citenamefont {Olovsson}\ \emph {et~al.}(2009)\citenamefont
  {Olovsson}, \citenamefont {Tanaka}, \citenamefont {Mizoguchi}, \citenamefont
  {Puschnig},\ and\ \citenamefont {Ambrosch-Draxl}}]{Olovsson2009}%
  \BibitemOpen
  \bibfield  {author} {\bibinfo {author} {\bibfnamefont {W.}~\bibnamefont
  {Olovsson}}, \bibinfo {author} {\bibfnamefont {I.}~\bibnamefont {Tanaka}},
  \bibinfo {author} {\bibfnamefont {T.}~\bibnamefont {Mizoguchi}}, \bibinfo
  {author} {\bibfnamefont {P.}~\bibnamefont {Puschnig}},\ and\ \bibinfo
  {author} {\bibfnamefont {C.}~\bibnamefont {Ambrosch-Draxl}},\ }\href
  {https://doi.org/10.1103/PhysRevB.79.041102} {\bibfield  {journal} {\bibinfo
  {journal} {Phys. Rev. B}\ }\textbf {\bibinfo {volume} {79}},\ \bibinfo
  {pages} {041102} (\bibinfo {year} {2009})}\BibitemShut {NoStop}%
\bibitem [{\citenamefont {Laskowski}\ and\ \citenamefont
  {Blaha}(2010)}]{Laskowski2010}%
  \BibitemOpen
  \bibfield  {author} {\bibinfo {author} {\bibfnamefont {R.}~\bibnamefont
  {Laskowski}}\ and\ \bibinfo {author} {\bibfnamefont {P.}~\bibnamefont
  {Blaha}},\ }\href {https://doi.org/10.1103/physrevb.82.205104} {\bibfield
  {journal} {\bibinfo  {journal} {Phys. Rev. B}\ }\textbf {\bibinfo {volume}
  {82}},\ \bibinfo {pages} {205104} (\bibinfo {year} {2010})}\BibitemShut
  {NoStop}%
\bibitem [{\citenamefont {Gao}\ \emph {et~al.}(2009)\citenamefont {Gao},
  \citenamefont {Pickard}, \citenamefont {Perlov},\ and\ \citenamefont
  {Milman}}]{Gao2009}%
  \BibitemOpen
  \bibfield  {author} {\bibinfo {author} {\bibfnamefont {S.-P.}\ \bibnamefont
  {Gao}}, \bibinfo {author} {\bibfnamefont {C.~J.}\ \bibnamefont {Pickard}},
  \bibinfo {author} {\bibfnamefont {A.}~\bibnamefont {Perlov}},\ and\ \bibinfo
  {author} {\bibfnamefont {V.}~\bibnamefont {Milman}},\ }\href
  {https://doi.org/10.1088/0953-8984/21/10/104203} {\bibfield  {journal}
  {\bibinfo  {journal} {J. Phys.: Condens. Matter}\ }\textbf {\bibinfo {volume}
  {21}},\ \bibinfo {pages} {104203} (\bibinfo {year} {2009})}\BibitemShut
  {NoStop}%
\bibitem [{\citenamefont {Gougoussis}\ \emph {et~al.}(2009)\citenamefont
  {Gougoussis}, \citenamefont {Calandra}, \citenamefont {Seitsonen},\ and\
  \citenamefont {Mauri}}]{Gougoussis2009}%
  \BibitemOpen
  \bibfield  {author} {\bibinfo {author} {\bibfnamefont {C.}~\bibnamefont
  {Gougoussis}}, \bibinfo {author} {\bibfnamefont {M.}~\bibnamefont
  {Calandra}}, \bibinfo {author} {\bibfnamefont {A.~P.}\ \bibnamefont
  {Seitsonen}},\ and\ \bibinfo {author} {\bibfnamefont {F.}~\bibnamefont
  {Mauri}},\ }\href {https://doi.org/10.1103/physrevb.80.075102} {\bibfield
  {journal} {\bibinfo  {journal} {Phys. Rev. B}\ }\textbf {\bibinfo {volume}
  {80}},\ \bibinfo {pages} {075102} (\bibinfo {year} {2009})}\BibitemShut
  {NoStop}%
\bibitem [{\citenamefont {Mazevet}\ \emph {et~al.}(2010)\citenamefont
  {Mazevet}, \citenamefont {Torrent}, \citenamefont {Recoules},\ and\
  \citenamefont {Jollet}}]{Mazevet2010}%
  \BibitemOpen
  \bibfield  {author} {\bibinfo {author} {\bibfnamefont {S.}~\bibnamefont
  {Mazevet}}, \bibinfo {author} {\bibfnamefont {M.}~\bibnamefont {Torrent}},
  \bibinfo {author} {\bibfnamefont {V.}~\bibnamefont {Recoules}},\ and\
  \bibinfo {author} {\bibfnamefont {F.}~\bibnamefont {Jollet}},\ }\href
  {https://doi.org/10.1016/j.hedp.2009.06.004} {\bibfield  {journal} {\bibinfo
  {journal} {High. Energ. Dens. Phys.}\ }\textbf {\bibinfo {volume} {6}},\
  \bibinfo {pages} {84} (\bibinfo {year} {2010})}\BibitemShut {NoStop}%
\bibitem [{\citenamefont {Bun{\u{a}}u}\ and\ \citenamefont
  {Calandra}(2013)}]{Bunau2013}%
  \BibitemOpen
  \bibfield  {author} {\bibinfo {author} {\bibfnamefont {O.}~\bibnamefont
  {Bun{\u{a}}u}}\ and\ \bibinfo {author} {\bibfnamefont {M.}~\bibnamefont
  {Calandra}},\ }\href {https://doi.org/10.1103/physrevb.87.205105} {\bibfield
  {journal} {\bibinfo  {journal} {Phys. Rev. B}\ }\textbf {\bibinfo {volume}
  {87}},\ \bibinfo {pages} {205105} (\bibinfo {year} {2013})}\BibitemShut
  {NoStop}%
\bibitem [{\citenamefont {Prentice}\ \emph {et~al.}(2020)\citenamefont
  {Prentice}, \citenamefont {Aarons}, \citenamefont {Womack}, \citenamefont
  {Allen}, \citenamefont {Andrinopoulos}, \citenamefont {Anton}, \citenamefont
  {Bell}, \citenamefont {Bhandari}, \citenamefont {Bramley}, \citenamefont
  {Charlton}, \citenamefont {Clements}, \citenamefont {Cole}, \citenamefont
  {Constantinescu}, \citenamefont {Corsetti}, \citenamefont {Dubois},
  \citenamefont {Duff}, \citenamefont {Escart{\'{\i}}n}, \citenamefont {Greco},
  \citenamefont {Hill}, \citenamefont {Lee}, \citenamefont {Linscott},
  \citenamefont {O'Regan}, \citenamefont {Phipps}, \citenamefont {Ratcliff},
  \citenamefont {Serrano}, \citenamefont {Tait}, \citenamefont {Teobaldi},
  \citenamefont {Vitale}, \citenamefont {Yeung}, \citenamefont {Zuehlsdorff},
  \citenamefont {Dziedzic}, \citenamefont {Haynes}, \citenamefont {Hine},
  \citenamefont {Mostofi}, \citenamefont {Payne},\ and\ \citenamefont
  {Skylaris}}]{Prentice2020}%
  \BibitemOpen
  \bibfield  {author} {\bibinfo {author} {\bibfnamefont {J.~C.~A.}\
  \bibnamefont {Prentice}}, \bibinfo {author} {\bibfnamefont {J.}~\bibnamefont
  {Aarons}}, \bibinfo {author} {\bibfnamefont {J.~C.}\ \bibnamefont {Womack}},
  \bibinfo {author} {\bibfnamefont {A.~E.~A.}\ \bibnamefont {Allen}}, \bibinfo
  {author} {\bibfnamefont {L.}~\bibnamefont {Andrinopoulos}}, \bibinfo {author}
  {\bibfnamefont {L.}~\bibnamefont {Anton}}, \bibinfo {author} {\bibfnamefont
  {R.~A.}\ \bibnamefont {Bell}}, \bibinfo {author} {\bibfnamefont
  {A.}~\bibnamefont {Bhandari}}, \bibinfo {author} {\bibfnamefont {G.~A.}\
  \bibnamefont {Bramley}}, \bibinfo {author} {\bibfnamefont {R.~J.}\
  \bibnamefont {Charlton}}, \bibinfo {author} {\bibfnamefont {R.~J.}\
  \bibnamefont {Clements}}, \bibinfo {author} {\bibfnamefont {D.~J.}\
  \bibnamefont {Cole}}, \bibinfo {author} {\bibfnamefont {G.}~\bibnamefont
  {Constantinescu}}, \bibinfo {author} {\bibfnamefont {F.}~\bibnamefont
  {Corsetti}}, \bibinfo {author} {\bibfnamefont {S.~M.-M.}\ \bibnamefont
  {Dubois}}, \bibinfo {author} {\bibfnamefont {K.~K.~B.}\ \bibnamefont {Duff}},
  \bibinfo {author} {\bibfnamefont {J.~M.}\ \bibnamefont {Escart{\'{\i}}n}},
  \bibinfo {author} {\bibfnamefont {A.}~\bibnamefont {Greco}}, \bibinfo
  {author} {\bibfnamefont {Q.}~\bibnamefont {Hill}}, \bibinfo {author}
  {\bibfnamefont {L.~P.}\ \bibnamefont {Lee}}, \bibinfo {author} {\bibfnamefont
  {E.}~\bibnamefont {Linscott}}, \bibinfo {author} {\bibfnamefont {D.~D.}\
  \bibnamefont {O'Regan}}, \bibinfo {author} {\bibfnamefont {M.~J.~S.}\
  \bibnamefont {Phipps}}, \bibinfo {author} {\bibfnamefont {L.~E.}\
  \bibnamefont {Ratcliff}}, \bibinfo {author} {\bibfnamefont {{\'{A}}.~R.}\
  \bibnamefont {Serrano}}, \bibinfo {author} {\bibfnamefont {E.~W.}\
  \bibnamefont {Tait}}, \bibinfo {author} {\bibfnamefont {G.}~\bibnamefont
  {Teobaldi}}, \bibinfo {author} {\bibfnamefont {V.}~\bibnamefont {Vitale}},
  \bibinfo {author} {\bibfnamefont {N.}~\bibnamefont {Yeung}}, \bibinfo
  {author} {\bibfnamefont {T.~J.}\ \bibnamefont {Zuehlsdorff}}, \bibinfo
  {author} {\bibfnamefont {J.}~\bibnamefont {Dziedzic}}, \bibinfo {author}
  {\bibfnamefont {P.~D.}\ \bibnamefont {Haynes}}, \bibinfo {author}
  {\bibfnamefont {N.~D.~M.}\ \bibnamefont {Hine}}, \bibinfo {author}
  {\bibfnamefont {A.~A.}\ \bibnamefont {Mostofi}}, \bibinfo {author}
  {\bibfnamefont {M.~C.}\ \bibnamefont {Payne}},\ and\ \bibinfo {author}
  {\bibfnamefont {C.-K.}\ \bibnamefont {Skylaris}},\ }\href
  {https://doi.org/10.1063/5.0004445} {\bibfield  {journal} {\bibinfo
  {journal} {J. Chem. Phys.}\ }\textbf {\bibinfo {volume} {152}},\ \bibinfo
  {pages} {174111} (\bibinfo {year} {2020})}\BibitemShut {NoStop}%
\bibitem [{\citenamefont {Karsai}\ \emph {et~al.}(2018)\citenamefont {Karsai},
  \citenamefont {Humer}, \citenamefont {Flage-Larsen}, \citenamefont {Blaha},\
  and\ \citenamefont {Kresse}}]{Karsai2018}%
  \BibitemOpen
  \bibfield  {author} {\bibinfo {author} {\bibfnamefont {F.}~\bibnamefont
  {Karsai}}, \bibinfo {author} {\bibfnamefont {M.}~\bibnamefont {Humer}},
  \bibinfo {author} {\bibfnamefont {E.}~\bibnamefont {Flage-Larsen}}, \bibinfo
  {author} {\bibfnamefont {P.}~\bibnamefont {Blaha}},\ and\ \bibinfo {author}
  {\bibfnamefont {G.}~\bibnamefont {Kresse}},\ }\href@noop {} {\bibfield
  {journal} {\bibinfo  {journal} {Phys. Rev. B}\ } (\bibinfo {year}
  {2018})}\BibitemShut {NoStop}%
\bibitem [{\citenamefont {Sander}\ \emph {et~al.}(2015)\citenamefont {Sander},
  \citenamefont {Maggio},\ and\ \citenamefont {Kresse}}]{Sander2015}%
  \BibitemOpen
  \bibfield  {author} {\bibinfo {author} {\bibfnamefont {T.}~\bibnamefont
  {Sander}}, \bibinfo {author} {\bibfnamefont {E.}~\bibnamefont {Maggio}},\
  and\ \bibinfo {author} {\bibfnamefont {G.}~\bibnamefont {Kresse}},\ }\href
  {https://doi.org/10.1103/PhysRevB.92.045209} {\bibfield  {journal} {\bibinfo
  {journal} {Phys. Rev. B}\ }\textbf {\bibinfo {volume} {92}},\ \bibinfo
  {pages} {045209} (\bibinfo {year} {2015})}\BibitemShut {NoStop}%
\bibitem [{\citenamefont {Hedin}(1965)}]{Hedin1965}%
  \BibitemOpen
  \bibfield  {author} {\bibinfo {author} {\bibfnamefont {L.}~\bibnamefont
  {Hedin}},\ }\href {https://doi.org/10.1103/PhysRev.139.A796} {\bibfield
  {journal} {\bibinfo  {journal} {Phys. Rev.}\ }\textbf {\bibinfo {volume}
  {139}},\ \bibinfo {pages} {A796} (\bibinfo {year} {1965})}\BibitemShut
  {NoStop}%
\bibitem [{\citenamefont {Strinati}(1988)}]{Strinati1988}%
  \BibitemOpen
  \bibfield  {author} {\bibinfo {author} {\bibfnamefont {G.}~\bibnamefont
  {Strinati}},\ }\href {https://doi.org/10.1007/bf02725962} {\bibfield
  {journal} {\bibinfo  {journal} {La Rivista del Nuovo Cimento}\ }\textbf
  {\bibinfo {volume} {11}},\ \bibinfo {pages} {1} (\bibinfo {year}
  {1988})}\BibitemShut {NoStop}%
\bibitem [{\citenamefont {Marini}\ and\ \citenamefont
  {Del~Sole}(2003)}]{Marini2003}%
  \BibitemOpen
  \bibfield  {author} {\bibinfo {author} {\bibfnamefont {A.}~\bibnamefont
  {Marini}}\ and\ \bibinfo {author} {\bibfnamefont {R.}~\bibnamefont
  {Del~Sole}},\ }\href {https://doi.org/10.1103/PhysRevLett.91.176402}
  {\bibfield  {journal} {\bibinfo  {journal} {Phys. Rev. Lett.}\ }\textbf
  {\bibinfo {volume} {91}},\ \bibinfo {pages} {176402} (\bibinfo {year}
  {2003})}\BibitemShut {NoStop}%
\bibitem [{\citenamefont {Stratmann}\ \emph {et~al.}(1998)\citenamefont
  {Stratmann}, \citenamefont {Scuseria},\ and\ \citenamefont
  {Frisch}}]{Stratmann1998}%
  \BibitemOpen
  \bibfield  {author} {\bibinfo {author} {\bibfnamefont {R.~E.}\ \bibnamefont
  {Stratmann}}, \bibinfo {author} {\bibfnamefont {G.~E.}\ \bibnamefont
  {Scuseria}},\ and\ \bibinfo {author} {\bibfnamefont {M.~J.}\ \bibnamefont
  {Frisch}},\ }\href {https://doi.org/10.1063/1.477483} {\bibfield  {journal}
  {\bibinfo  {journal} {J. Chem. Phys.}\ }\textbf {\bibinfo {volume} {109}},\
  \bibinfo {pages} {8218} (\bibinfo {year} {1998})}\BibitemShut {NoStop}%
\bibitem [{\citenamefont {Furche}(2001)}]{Furche2001}%
  \BibitemOpen
  \bibfield  {author} {\bibinfo {author} {\bibfnamefont {F.}~\bibnamefont
  {Furche}},\ }\href {https://doi.org/10.1103/physrevb.64.195120} {\bibfield
  {journal} {\bibinfo  {journal} {Phys. Rev. B}\ }\textbf {\bibinfo {volume}
  {64}},\ \bibinfo {pages} {195120} (\bibinfo {year} {2001})}\BibitemShut
  {NoStop}%
\bibitem [{\citenamefont {Dancoff}(1950)}]{Dancoff1950}%
  \BibitemOpen
  \bibfield  {author} {\bibinfo {author} {\bibfnamefont {S.~M.}\ \bibnamefont
  {Dancoff}},\ }\href {https://doi.org/10.1103/physrev.78.382} {\bibfield
  {journal} {\bibinfo  {journal} {Phys. Rev.}\ }\textbf {\bibinfo {volume}
  {78}},\ \bibinfo {pages} {382} (\bibinfo {year} {1950})}\BibitemShut
  {NoStop}%
\bibitem [{\citenamefont {Tamm}(1991)}]{Tamm1991}%
  \BibitemOpen
  \bibfield  {author} {\bibinfo {author} {\bibfnamefont {I.~E.}\ \bibnamefont
  {Tamm}},\ }\href {https://doi.org/10.1007/978-3-642-74626-0} {\emph {\bibinfo
  {title} {Selected Papers}}},\ edited by\ \bibinfo {editor} {\bibfnamefont
  {B.~M.}\ \bibnamefont {Bolotovskii}}, \bibinfo {editor} {\bibfnamefont
  {V.~Y.}\ \bibnamefont {Frenkel}},\ and\ \bibinfo {editor} {\bibfnamefont
  {R.}~\bibnamefont {Peierls}}\ (\bibinfo  {publisher} {Springer Berlin
  Heidelberg},\ \bibinfo {year} {1991})\BibitemShut {NoStop}%
\bibitem [{\citenamefont {Blöchl}(1994)}]{Bloechl1994}%
  \BibitemOpen
  \bibfield  {author} {\bibinfo {author} {\bibfnamefont {P.~E.}\ \bibnamefont
  {Blöchl}},\ }\href {https://doi.org/10.1103/physrevb.50.17953} {\bibfield
  {journal} {\bibinfo  {journal} {Phys. Rev. B}\ }\textbf {\bibinfo {volume}
  {50}},\ \bibinfo {pages} {17953} (\bibinfo {year} {1994})}\BibitemShut
  {NoStop}%
\bibitem [{\citenamefont {Kresse}\ and\ \citenamefont
  {Joubert}(1999)}]{Kresse1999}%
  \BibitemOpen
  \bibfield  {author} {\bibinfo {author} {\bibfnamefont {G.}~\bibnamefont
  {Kresse}}\ and\ \bibinfo {author} {\bibfnamefont {D.}~\bibnamefont
  {Joubert}},\ }\href {https://doi.org/10.1103/physrevb.59.1758} {\bibfield
  {journal} {\bibinfo  {journal} {Phys. Rev. B}\ }\textbf {\bibinfo {volume}
  {59}},\ \bibinfo {pages} {1758} (\bibinfo {year} {1999})}\BibitemShut
  {NoStop}%
\bibitem [{\citenamefont {Kresse}\ and\ \citenamefont
  {Furthmüller}(1996)}]{Kresse1996}%
  \BibitemOpen
  \bibfield  {author} {\bibinfo {author} {\bibfnamefont {G.}~\bibnamefont
  {Kresse}}\ and\ \bibinfo {author} {\bibfnamefont {J.}~\bibnamefont
  {Furthmüller}},\ }\href {https://doi.org/10.1016/0927-0256(96)00008-0}
  {\bibfield  {journal} {\bibinfo  {journal} {Comput. Mater. Sci.}\ }\textbf
  {\bibinfo {volume} {6}},\ \bibinfo {pages} {15} (\bibinfo {year}
  {1996})}\BibitemShut {NoStop}%
\bibitem [{\citenamefont {Perdew}\ \emph {et~al.}(1996)\citenamefont {Perdew},
  \citenamefont {Burke},\ and\ \citenamefont {Ernzerhof}}]{Perdew1996}%
  \BibitemOpen
  \bibfield  {author} {\bibinfo {author} {\bibfnamefont {J.~P.}\ \bibnamefont
  {Perdew}}, \bibinfo {author} {\bibfnamefont {K.}~\bibnamefont {Burke}},\ and\
  \bibinfo {author} {\bibfnamefont {M.}~\bibnamefont {Ernzerhof}},\ }\href
  {https://doi.org/10.1103/physrevlett.77.3865} {\bibfield  {journal} {\bibinfo
   {journal} {Phys. Rev. Lett.}\ }\textbf {\bibinfo {volume} {77}},\ \bibinfo
  {pages} {3865} (\bibinfo {year} {1996})}\BibitemShut {NoStop}%
\bibitem [{\citenamefont {Brandes}\ \emph {et~al.}(2008)\citenamefont
  {Brandes}, \citenamefont {Cody}, \citenamefont {Rumble}, \citenamefont
  {Haberstroh}, \citenamefont {Wirick},\ and\ \citenamefont
  {Gelinas}}]{Brandes2008}%
  \BibitemOpen
  \bibfield  {author} {\bibinfo {author} {\bibfnamefont {J.~A.}\ \bibnamefont
  {Brandes}}, \bibinfo {author} {\bibfnamefont {G.~D.}\ \bibnamefont {Cody}},
  \bibinfo {author} {\bibfnamefont {D.}~\bibnamefont {Rumble}}, \bibinfo
  {author} {\bibfnamefont {P.}~\bibnamefont {Haberstroh}}, \bibinfo {author}
  {\bibfnamefont {S.}~\bibnamefont {Wirick}},\ and\ \bibinfo {author}
  {\bibfnamefont {Y.}~\bibnamefont {Gelinas}},\ }\href
  {https://doi.org/10.1016/j.carbon.2008.06.020} {\bibfield  {journal}
  {\bibinfo  {journal} {Carbon}\ }\textbf {\bibinfo {volume} {46}},\ \bibinfo
  {pages} {1424} (\bibinfo {year} {2008})}\BibitemShut {NoStop}%
\bibitem [{\citenamefont {Li}\ \emph {et~al.}(2012)\citenamefont {Li},
  \citenamefont {Petravic}, \citenamefont {Cowie}, \citenamefont {Xing},
  \citenamefont {Peter}, \citenamefont {Chen}, \citenamefont {Si},\ and\
  \citenamefont {Duan}}]{Li2012}%
  \BibitemOpen
  \bibfield  {author} {\bibinfo {author} {\bibfnamefont {L.~H.}\ \bibnamefont
  {Li}}, \bibinfo {author} {\bibfnamefont {M.}~\bibnamefont {Petravic}},
  \bibinfo {author} {\bibfnamefont {B.~C.~C.}\ \bibnamefont {Cowie}}, \bibinfo
  {author} {\bibfnamefont {T.}~\bibnamefont {Xing}}, \bibinfo {author}
  {\bibfnamefont {R.}~\bibnamefont {Peter}}, \bibinfo {author} {\bibfnamefont
  {Y.}~\bibnamefont {Chen}}, \bibinfo {author} {\bibfnamefont {C.}~\bibnamefont
  {Si}},\ and\ \bibinfo {author} {\bibfnamefont {W.}~\bibnamefont {Duan}},\
  }\href {https://doi.org/10.1063/1.4767135} {\bibfield  {journal} {\bibinfo
  {journal} {Appl. Phys. Lett.}\ }\textbf {\bibinfo {volume} {101}},\ \bibinfo
  {pages} {191604} (\bibinfo {year} {2012})}\BibitemShut {NoStop}%
\bibitem [{\citenamefont {Petravic}\ \emph {et~al.}(2013)\citenamefont
  {Petravic}, \citenamefont {Peter}, \citenamefont {Varasanec}, \citenamefont
  {Li}, \citenamefont {Chen},\ and\ \citenamefont {Cowie}}]{Petravic2013}%
  \BibitemOpen
  \bibfield  {author} {\bibinfo {author} {\bibfnamefont {M.}~\bibnamefont
  {Petravic}}, \bibinfo {author} {\bibfnamefont {R.}~\bibnamefont {Peter}},
  \bibinfo {author} {\bibfnamefont {M.}~\bibnamefont {Varasanec}}, \bibinfo
  {author} {\bibfnamefont {L.~H.}\ \bibnamefont {Li}}, \bibinfo {author}
  {\bibfnamefont {Y.}~\bibnamefont {Chen}},\ and\ \bibinfo {author}
  {\bibfnamefont {B.~C.~C.}\ \bibnamefont {Cowie}},\ }\href
  {https://doi.org/10.1116/1.4798271} {\bibfield  {journal} {\bibinfo
  {journal} {Journal of Vacuum Science {\&} Technology A: Vacuum, Surfaces, and
  Films}\ }\textbf {\bibinfo {volume} {31}},\ \bibinfo {pages} {031405}
  (\bibinfo {year} {2013})}\BibitemShut {NoStop}%
\bibitem [{\citenamefont {Liu}\ \emph {et~al.}(2016)\citenamefont {Liu},
  \citenamefont {Kaltak}, \citenamefont {Klimeš},\ and\ \citenamefont
  {Kresse}}]{Liu2016}%
  \BibitemOpen
  \bibfield  {author} {\bibinfo {author} {\bibfnamefont {P.}~\bibnamefont
  {Liu}}, \bibinfo {author} {\bibfnamefont {M.}~\bibnamefont {Kaltak}},
  \bibinfo {author} {\bibfnamefont {J.}~\bibnamefont {Klimeš}},\ and\ \bibinfo
  {author} {\bibfnamefont {G.}~\bibnamefont {Kresse}},\ }\href
  {https://doi.org/10.1103/PhysRevB.94.165109} {\bibfield  {journal} {\bibinfo
  {journal} {Phys. Rev. B}\ }\textbf {\bibinfo {volume} {94}},\ \bibinfo
  {pages} {165109} (\bibinfo {year} {2016})}\BibitemShut {NoStop}%
\bibitem [{\citenamefont {Bokdam}\ \emph {et~al.}(2016)\citenamefont {Bokdam},
  \citenamefont {Sander}, \citenamefont {Stroppa}, \citenamefont {Picozzi},
  \citenamefont {Sarma}, \citenamefont {Franchini},\ and\ \citenamefont
  {Kresse}}]{Bokdam2016}%
  \BibitemOpen
  \bibfield  {author} {\bibinfo {author} {\bibfnamefont {M.}~\bibnamefont
  {Bokdam}}, \bibinfo {author} {\bibfnamefont {T.}~\bibnamefont {Sander}},
  \bibinfo {author} {\bibfnamefont {A.}~\bibnamefont {Stroppa}}, \bibinfo
  {author} {\bibfnamefont {S.}~\bibnamefont {Picozzi}}, \bibinfo {author}
  {\bibfnamefont {D.~D.}\ \bibnamefont {Sarma}}, \bibinfo {author}
  {\bibfnamefont {C.}~\bibnamefont {Franchini}},\ and\ \bibinfo {author}
  {\bibfnamefont {G.}~\bibnamefont {Kresse}},\ }\href@noop {} {\bibfield
  {journal} {\bibinfo  {journal} {Sci. Rep.}\ }\textbf {\bibinfo {volume} {6}}
  (\bibinfo {year} {2016})}\BibitemShut {NoStop}%
\bibitem [{\citenamefont {Ma}\ \emph {et~al.}(1992)\citenamefont {Ma},
  \citenamefont {Wassdahl}, \citenamefont {Skytt}, \citenamefont {Guo},
  \citenamefont {Nordgren}, \citenamefont {Johnson}, \citenamefont {Rubensson},
  \citenamefont {Boske}, \citenamefont {Eberhardt},\ and\ \citenamefont
  {Kevan}}]{Ma1992}%
  \BibitemOpen
  \bibfield  {author} {\bibinfo {author} {\bibfnamefont {Y.}~\bibnamefont
  {Ma}}, \bibinfo {author} {\bibfnamefont {N.}~\bibnamefont {Wassdahl}},
  \bibinfo {author} {\bibfnamefont {P.}~\bibnamefont {Skytt}}, \bibinfo
  {author} {\bibfnamefont {J.}~\bibnamefont {Guo}}, \bibinfo {author}
  {\bibfnamefont {J.}~\bibnamefont {Nordgren}}, \bibinfo {author}
  {\bibfnamefont {P.~D.}\ \bibnamefont {Johnson}}, \bibinfo {author}
  {\bibfnamefont {J.-E.}\ \bibnamefont {Rubensson}}, \bibinfo {author}
  {\bibfnamefont {T.}~\bibnamefont {Boske}}, \bibinfo {author} {\bibfnamefont
  {W.}~\bibnamefont {Eberhardt}},\ and\ \bibinfo {author} {\bibfnamefont
  {S.~D.}\ \bibnamefont {Kevan}},\ }\href
  {https://doi.org/10.1103/physrevlett.69.2598} {\bibfield  {journal} {\bibinfo
   {journal} {Phys. Rev. Lett.}\ }\textbf {\bibinfo {volume} {69}},\ \bibinfo
  {pages} {2598} (\bibinfo {year} {1992})}\BibitemShut {NoStop}%
\bibitem [{\citenamefont {Shirley}(2000)}]{Shirley2000}%
  \BibitemOpen
  \bibfield  {author} {\bibinfo {author} {\bibfnamefont {E.~L.}\ \bibnamefont
  {Shirley}},\ }\href {https://doi.org/10.1016/s0368-2048(00)00170-5}
  {\bibfield  {journal} {\bibinfo  {journal} {J. Electron Spectrosc. Relat.
  Phenom.}\ }\textbf {\bibinfo {volume} {110-111}},\ \bibinfo {pages} {305}
  (\bibinfo {year} {2000})}\BibitemShut {NoStop}%
\bibitem [{\citenamefont {Olovsson}\ \emph {et~al.}(2019)\citenamefont
  {Olovsson}, \citenamefont {Mizoguchi}, \citenamefont {Magnuson},
  \citenamefont {Kontur}, \citenamefont {Hellman}, \citenamefont {Tanaka},\
  and\ \citenamefont {Draxl}}]{Olovsson2019}%
  \BibitemOpen
  \bibfield  {author} {\bibinfo {author} {\bibfnamefont {W.}~\bibnamefont
  {Olovsson}}, \bibinfo {author} {\bibfnamefont {T.}~\bibnamefont {Mizoguchi}},
  \bibinfo {author} {\bibfnamefont {M.}~\bibnamefont {Magnuson}}, \bibinfo
  {author} {\bibfnamefont {S.}~\bibnamefont {Kontur}}, \bibinfo {author}
  {\bibfnamefont {O.}~\bibnamefont {Hellman}}, \bibinfo {author} {\bibfnamefont
  {I.}~\bibnamefont {Tanaka}},\ and\ \bibinfo {author} {\bibfnamefont
  {C.}~\bibnamefont {Draxl}},\ }\href@noop {} {\bibfield  {journal} {\bibinfo
  {journal} {The Journal of Physical Chemistry C}\ }\textbf {\bibinfo {volume}
  {123}},\ \bibinfo {pages} {9688} (\bibinfo {year} {2019})}\BibitemShut
  {NoStop}%
\bibitem [{\citenamefont {Brühwiler}\ \emph {et~al.}(1995)\citenamefont
  {Brühwiler}, \citenamefont {Maxwell}, \citenamefont {Puglia}, \citenamefont
  {Nilsson}, \citenamefont {Andersson},\ and\ \citenamefont
  {M{\aa}rtensson}}]{Bruehwiler1995}%
  \BibitemOpen
  \bibfield  {author} {\bibinfo {author} {\bibfnamefont {P.~A.}\ \bibnamefont
  {Brühwiler}}, \bibinfo {author} {\bibfnamefont {A.~J.}\ \bibnamefont
  {Maxwell}}, \bibinfo {author} {\bibfnamefont {C.}~\bibnamefont {Puglia}},
  \bibinfo {author} {\bibfnamefont {A.}~\bibnamefont {Nilsson}}, \bibinfo
  {author} {\bibfnamefont {S.}~\bibnamefont {Andersson}},\ and\ \bibinfo
  {author} {\bibfnamefont {N.}~\bibnamefont {M{\aa}rtensson}},\ }\href
  {https://doi.org/10.1103/physrevlett.74.614} {\bibfield  {journal} {\bibinfo
  {journal} {Phys. Rev. Lett.}\ }\textbf {\bibinfo {volume} {74}},\ \bibinfo
  {pages} {614} (\bibinfo {year} {1995})}\BibitemShut {NoStop}%
\bibitem [{\citenamefont {McDougall}\ \emph {et~al.}(2014)\citenamefont
  {McDougall}, \citenamefont {Nicholls}, \citenamefont {Partridge},\ and\
  \citenamefont {McCulloch}}]{McDougall2014}%
  \BibitemOpen
  \bibfield  {author} {\bibinfo {author} {\bibfnamefont {N.~L.}\ \bibnamefont
  {McDougall}}, \bibinfo {author} {\bibfnamefont {R.~J.}\ \bibnamefont
  {Nicholls}}, \bibinfo {author} {\bibfnamefont {J.~G.}\ \bibnamefont
  {Partridge}},\ and\ \bibinfo {author} {\bibfnamefont {D.~G.}\ \bibnamefont
  {McCulloch}},\ }\href {https://doi.org/10.1017/s1431927614000737} {\bibfield
  {journal} {\bibinfo  {journal} {Microsc. Microanal.}\ }\textbf {\bibinfo
  {volume} {20}},\ \bibinfo {pages} {1053} (\bibinfo {year}
  {2014})}\BibitemShut {NoStop}%
\bibitem [{\citenamefont {Handa}\ \emph {et~al.}(2005)\citenamefont {Handa},
  \citenamefont {Kojima}, \citenamefont {Ozutsumi}, \citenamefont {Taniguchi},\
  and\ \citenamefont {Ikeda}}]{Handa2005}%
  \BibitemOpen
  \bibfield  {author} {\bibinfo {author} {\bibfnamefont {K.}~\bibnamefont
  {Handa}}, \bibinfo {author} {\bibfnamefont {K.}~\bibnamefont {Kojima}},
  \bibinfo {author} {\bibfnamefont {K.}~\bibnamefont {Ozutsumi}}, \bibinfo
  {author} {\bibfnamefont {K.}~\bibnamefont {Taniguchi}},\ and\ \bibinfo
  {author} {\bibfnamefont {S.}~\bibnamefont {Ikeda}},\ }\href@noop {}
  {\bibfield  {journal} {\bibinfo  {journal} {Memoirs Of The SR Center
  Ritsumeikan University}\ }\textbf {\bibinfo {volume} {07}} (\bibinfo {year}
  {2005})}\BibitemShut {NoStop}%
\bibitem [{\citenamefont {Het{\'{e}}nyi}\ \emph {et~al.}(2004)\citenamefont
  {Het{\'{e}}nyi}, \citenamefont {Angelis}, \citenamefont {Giannozzi},\ and\
  \citenamefont {Car}}]{Hetenyi2004}%
  \BibitemOpen
  \bibfield  {author} {\bibinfo {author} {\bibfnamefont {B.}~\bibnamefont
  {Het{\'{e}}nyi}}, \bibinfo {author} {\bibfnamefont {F.~D.}\ \bibnamefont
  {Angelis}}, \bibinfo {author} {\bibfnamefont {P.}~\bibnamefont {Giannozzi}},\
  and\ \bibinfo {author} {\bibfnamefont {R.}~\bibnamefont {Car}},\ }\href
  {https://doi.org/10.1063/1.1703526} {\bibfield  {journal} {\bibinfo
  {journal} {J. Chem. Phys.}\ }\textbf {\bibinfo {volume} {120}},\ \bibinfo
  {pages} {8632} (\bibinfo {year} {2004})}\BibitemShut {NoStop}%
\bibitem [{\citenamefont {Prendergast}\ and\ \citenamefont
  {Galli}(2006)}]{Prendergast2006}%
  \BibitemOpen
  \bibfield  {author} {\bibinfo {author} {\bibfnamefont {D.}~\bibnamefont
  {Prendergast}}\ and\ \bibinfo {author} {\bibfnamefont {G.}~\bibnamefont
  {Galli}},\ }\href {https://doi.org/10.1103/physrevlett.96.215502} {\bibfield
  {journal} {\bibinfo  {journal} {Phys. Rev. Lett.}\ }\textbf {\bibinfo
  {volume} {96}},\ \bibinfo {pages} {215502} (\bibinfo {year}
  {2006})}\BibitemShut {NoStop}%
\end{thebibliography}%

\end{document}